\documentclass[10pt]{iopart}

\usepackage{graphicx}
\usepackage[labelfont=bf,font=small]{caption}
\usepackage{amssymb}

\usepackage{color}

\begin{document}

\title[BH-NS merger ejecta]{Dynamics, nucleosynthesis, and kilonova signature
of black hole - neutron star merger ejecta}

\author{Rodrigo Fern\'andez$^{1,2,3}$, Francois Foucart$^{4,5}$, \\Daniel Kasen$^{2,3,4}$, 
        Jonas Lippuner$^6$, Dhruv Desai$^2$, \\Luke F. Roberts$^7$}
\address{$^1$ Department of Physics, University of Alberta, Edmonton, AB T6G 2E1, Canada.}
\address{$^2$ Department of Physics, University of California, Berkeley, CA 94720, USA.}
\address{$^3$ Department of Astronomy \& Theoretical Astrophysics Center, 
              University of California, Berkeley, CA 94720, USA.}
\address{$^4$ Nuclear Science Division, Lawrence Berkeley National Laboratory, Berkeley, CA 94720, USA.}
\address{$^5$ NASA Einstein Fellow}
\address{$^6$ TAPIR, Walter Burke Institute for Theoretical Physics, California Institute of Technology,
              Pasadena, CA 99125, USA.}
\address{$^7$ National Superconducting Cyclotron Laboratory and Department of Physics and Astronomy, 
	      Michigan State University, East Lansing, MI 48824, USA}
\ead{rafernan@ualberta.ca}

\begin{abstract}
We investigate the ejecta from black hole - neutron star mergers
by modeling the formation and interaction of mass ejected in a tidal 
tail and a disk wind. The outflows are
neutron-rich, giving rise to optical/infrared emission powered 
by the radioactive decay of $r$-process elements (a \emph{kilonova}). Here we perform
an end-to-end study of this phenomenon, 
where we start from the output of a fully-relativistic merger simulation, 
calculate the post-merger hydrodynamical evolution of the ejecta and disk winds 
including neutrino physics, determine the final nucleosynthetic yields using
post-processing nuclear reaction network calculations, and compute the kilonova emission 
with a radiative transfer code.
We study the effects of the tail-to-disk mass ratio by
scaling the tail density.
A larger initial tail mass results in fallback matter
becoming mixed into the disk and ejected in the subsequent disk wind.
Relative to the case of a disk without dynamical ejecta, the combined outflow has lower mean
electron fraction, faster speed, larger total mass, and larger absolute mass free of 
high-opacity Lanthanides or Actinides.
In most cases, the nucleosynthetic yield is dominated by the heavy $r$-process contribution from the
unbound part of the dynamical ejecta. A Solar-like abundance distribution 
can however be obtained when the total  mass of the dynamical ejecta 
is comparable to the mass of the disk outflows.
The kilonova has a characteristic duration of 1 week and a luminosity of $\sim
10^{41}$~erg~s$^{-1}$, with orientation effects leading to variations of a
factor $\sim 2$ in brightness. At early times ($< 1$~day) the emission includes
an optical component from the (hot) Lanthanide-rich material, but the spectrum
evolves quickly to the infrared thereafter.
\end{abstract}

\vspace{2pc}
\noindent{\it Keywords}: accretion, accretion disks -- dense matter -- gravitational waves
                       -- hydrodynamics -- neutrinos -- nuclear reactions, nucleosynthesis, abundances.

\section{Introduction}

The recent success of Advanced LIGO in detecting gravitational
waves (GWs) from binary black hole (BH-BH) mergers \cite{gw150914_prl,gw151226_prl} 
has marked the onset of GW Astronomy. As the sensitivity
of detectors improves, detection of double neutron star (NS-NS)
and NS-BH mergers in GWs is expected within the next few years \cite{Abadie+10}. 
Mergers that involve NSs are of interest to a broad community because they
can help to test the equation of state (EOS) of dense matter (e.g., \cite{Bauswein+13_prl}), 
are a prime candidate astrophysical site for $r$-process elements 
(e.g., \cite{just2014,roberts2016,wu2016}), and are expected to produce 
electromagnetic (EM) counterparts over a wide range of timescales and 
wavelengths (e.g., \cite{rosswog2015,FM16}).

The most easily detectable EM counterpart of NS-NS/NS-BH mergers
is a supernova-like transient powered by the radioactive decay
of $r$-process elements produced in the expanding ejecta, commonly
known as a \emph{kilonova} or \emph{macronova}
\cite{Li&Pacyznski98,Metzger+10b,Roberts+11,tanaka2016,metzger2016}. These transients 
arise from the sub-relativistic ejecta of the merger and are hence
not affected by beaming like short gamma-ray bursts (SGRBs). The emission
is detectable at optical and infrared wavelengths \cite{Metzger&Berger12}.

The intimate relation between a kilonova and the production of $r$-process elements
makes the transient a powerful diagnostic of the physical conditions
in the merger. This diagnostic power arises from the sensitivity of the 
optical opacity to the type of $r$-process composition of the ejecta:
even a small fraction of Lanthanides or Actinides
(mass number $A \gtrsim 140$) can increase the optical opacity by
orders of magnitude relative to iron-group-like composition \cite{kasen2013,fontes2015}.
The resulting kilonova can range from an optical transient lasting
days, to an infrared transient lasting for weeks \cite{barnes2013,tanaka2013,hotokezaka2016,
barnes2016,rosswog2016}.

Two distinct sources of merger ejecta with potentially different compositions
contribute to the kilonova emission. 
Material expelled on a dynamical time
leaves the system first, with typical velocities $0.1-0.3c$ (e.g., \cite{Hotokezaka+13}). 
For NS-BH mergers, ejection is driven mainly by tidal forces, and therefore
the ejecta is launched primarily on the equatorial plane. For NS-NS mergers
additional ejection occurs from the contact interface, leading 
to an outflow more widely distributed in solid angle. The tidally ejected material is mostly neutron-rich, generating a 
robust abundance of elements with $A > 130$ that follow the Solar System $r$-process 
distribution \cite{Korobkin+12,bauswein2013}. A shock-heated and/or neutrino-irradiated
component of the interface ejecta can have high electron fraction $Y_e$, by virtue of weak interactions, 
thereby producing lighter $r$-process elements \cite{wanajo2014}. 
The prevalence of the high-$Y_e$ component is an open research question, however, as it is sensitive to
the treatment of neutrinos and the nuclear EOS 
\cite{sekiguchi2015,foucart2015,palenzuela_2015,roberts2016,lehner_2016,radice2016,foucart2016}.

The second source of ejecta is the remnant accretion disk. Outflows
can be launched on the thermal time ($\sim 10-100$~ms) if there is sufficient neutrino
irradiation from a hypermassive NS \cite{Ruffert+97,Dessart+09,perego2014}, or on longer 
times $\gtrsim 1$~s once neutrino emission has subsided and the disk reaches the advective 
state \cite{metzger09b,Lee+09,FM13,just2014}. 
Magnetically-driven disk winds can also contribute to these disk outflows at
early times during the transient phase (e.g.~\cite{Kiuchi:2015qua}) or in the
late disk evolution. The disk ejecta component is slower (velocity $\sim 0.05$c) than the
dynamical ejecta, and is generally less neutron-rich due to longer
exposure to weak interactions. The resulting composition can be dominated
by light $r$-process elements. Whether heavier elements ($A>130$) are also produced 
depends on astrophysical parameters such as the disk mass or the properties of
angular momentum transport \cite{just2014,wu2016}. The amount of mass ejected
can be comparable or even dominate the dynamical ejecta, depending
on the binary parameters (e.g., \cite{FM16}).

The different compositions and kinematic properties of these two ejecta
channels have implications for the magnitude, color, and duration of the kilonova: 
opacity and velocity control the diffusion time. Furthermore, the expected geometry of 
NS-BH merger ejecta is such that the kilonova color can have a strong viewing angle 
dependency \cite{Kasen+15}. Likewise, the dynamical ejecta has a gravitationally bound 
component that falls back onto the central object, potentially interacting with the disk outflow 
and hence altering the net composition and mass ejection from the disk. 
A first attempt at characterizing the interplay between these two ejecta 
components was made by \cite{FQSKR2015}
for the case of NS-BH mergers. Starting from the output of a Newtonian merger
simulation, the long-term evolution of all components was followed until homology was reached.
The disk outflow was found to suppress fallback accretion relative to the case in
which the disk is absent, with long-term engine activity still possible
by accretion from the disk. No significant changes in the disk outflow properties 
were found relative to the case without dynamical ejecta, although only
one realization was studied in which the disk has a high mass ($0.2M_\odot$) relative
to the bound dynamical ejecta ($0.02M_\odot$).

Here we improve upon the work of \cite{FQSKR2015} by using initial conditions
mapped from a fully relativistic NS-BH simulation that includes
neutrino irradiation and therefore provides more realistic
initial conditions.
In addition, we explore the effect of varying the relative initial masses of 
dynamical ejecta and disk, by scaling the initial dynamical ejecta density. 
Given that the dynamical ejecta moves nearly ballistically, this is a good first approximation
to test the impact the ratio of dynamical- to disk ejecta without having to perform a large number of costly
merger simulations. 
We conduct the post-merger simulations in axisymmetry, and include the dominant energy and lepton 
number source terms, in addition to passive tracer particles. The output is post-processed 
with a nuclear reaction network and a radiative transfer code to compute $r$-process 
nucleosynthesis yields and kilonova light curve and spectral predictions, respectively.

The structure of this paper is the following. Section 2 describes our computational
method, Section 3 presents the results of our hydrodynamic simulations of the remnant
evolution, Section 4 describes our nucleosynthesis results, Section 5 our
kilonova predictions, and Section 6 presents our summary and discussion.

\section{Methods}

\subsection{Initial conditions}
\label{s:spec}

\begin{figure*}
\includegraphics*[width=\textwidth]{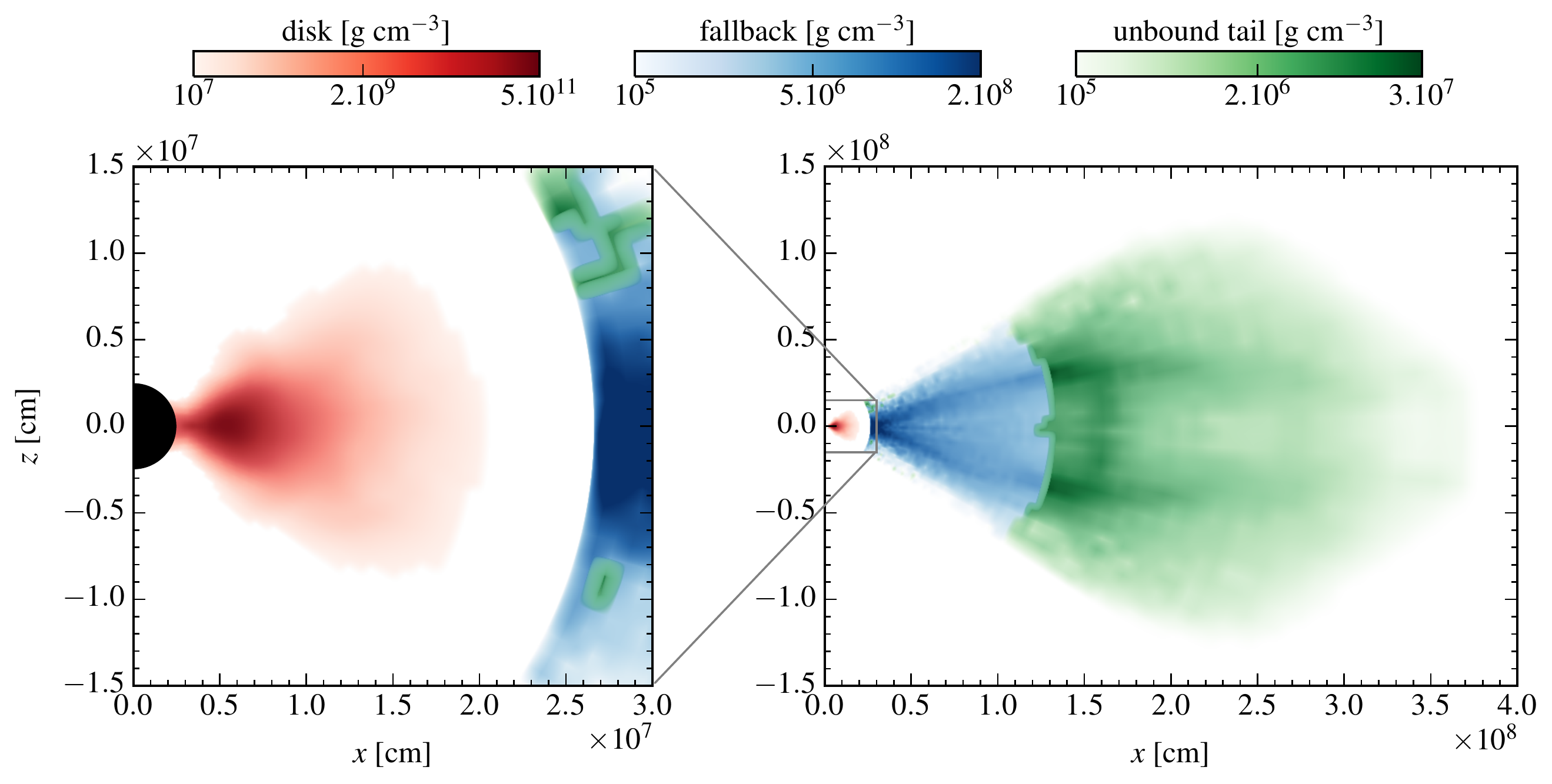}
\includegraphics*[width=\textwidth]{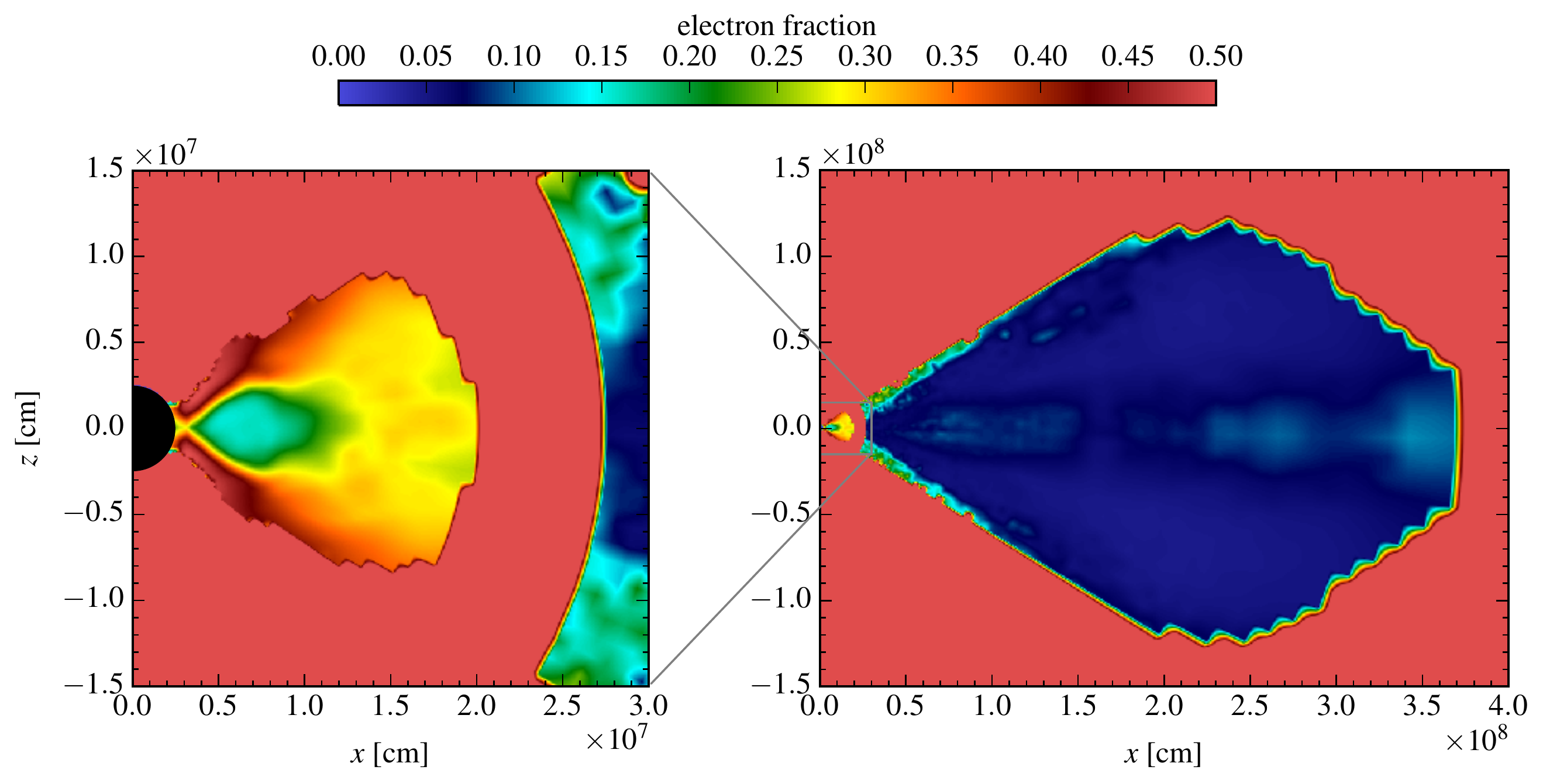}
\caption{\emph{Top:} Partial densities ($=\rho X_i$, with $X_i$ the mass fraction) 
of disk, bound dynamical ejecta (``fallback"), and unbound dynamical ejecta (``unbound tail") components at 
$t=0$ for the baseline model F0. Note that the color scale is not the same for all
components; here it is chosen to maximize contrast. \emph{Bottom:} Electron fraction at
$t=0$ in model F0. The dynamical ejecta density is reconstructed from tracer particles and
re-mapped onto the spherical grid. Rough edges are smoothed out by evolving the entire system for 
$\sim 10^{-4}$~s without source terms. The gap between disk  and dynamical ejecta -- introduced to eliminate 
overlap between the disk distribution and dynamical ejecta distribution -- is also filled within that timescale.}
\label{f:ic}
\end{figure*}

The initial condition for our models is the final
snapshot of a BH-NS post-merger remnant simulation presented in 
\cite{foucart2015}. This simulation evolves
the accretion disk resulting from the merger of a $7M_\odot$ BH with a $1.4M_\odot$
NS (case M14-7-S8 from~\cite{foucart2014}), up to $20\,{\rm ms}$ after the merger. 
The dimensionless spin of the BH before merger is $\chi=0.8$,
aligned with the orbital angular momentum of the system. After merger, 
the BH spin is $\chi=0.861$. 
The inspiral, merger, and post-merger evolution are performed using 
the SpEC code~\cite{SpECwebsite}, which evolves Einstein's equations of general relativity
in the Generalized Harmonic formalism using pseudospectral methods~\cite{Lindblom2006}, 
the relativistic equations of hydrodynamics in conservative form using high-order shock
capturing finite volume methods~\cite{Duez:2008rb,Foucart:2013a}, and the first two moments
of the neutrino distribution function with an analytical M1 closure~\cite{thorne80,shibata:11,foucart2015}.
The equation of state for the neutron star is that of
Lattimer and Swesty (1991)~\cite{Lattimer:1991nc}, using a compressibility parameter
$K_0=220\,{\rm MeV}$.
The computational domain in SpEC encompasses the
central object and the accretion disk, while dynamical ejecta with a fallback time longer
than $\sim 20\,{\rm ms}$
is allowed to leave the domain. 
Up to the disruption of the neutron star by the gravitational field of the black hole,
the SpEC evolution explicitly imposes symmetry of the fluid variables across the equatorial
plane, while the disk formation and post-merger evolution is performed without that symmetry
requirement (although SpEC preserves quite accurately the original symmetry of the system).
We note that while for the system studied here the exact solution is symmetric across the equatorial
plane, explicitly imposing this condition in simulations can create some numerical artifacts in the 
results. In particular, small errors in the vertical component of the fluid velocity in the
equatorial plane can push material away from that plane, leading to density
maxima offset from the equator. While this effect decreases as the numerical resolution increases,
it can be easily observed in lower-resolution regions - e.g. the low density regions of the tidal
tail, shown in Fig.~\ref{f:ic}.

We generate initial conditions for our
2D models by axisymmetrizing the matter distribution from the SpEC
simulation, keeping only the data with density higher than $10^8$~g~cm$^{-3}$, as
lower density material is not reliably evolved in the general relativistic merger
simulation. The axisymmetric baryon density $\tilde \rho$ is obtained
by integrating the baryon density $\rho$ over a curve $L(r,z)$ of constant coordinate height $z$
and constant cylindrical radius $r$, and then dividing by the proper length of $L$. 
For other axisymmetric quantities $\tilde X$, we use density-weighted averages, i.e.
\begin{equation}
\tilde X(r,z) = \frac{\oint_{L(r,z)} X \rho W \sqrt{g} dl}{\oint_{L(r,z)}  \rho W \sqrt{g} dl}
\end{equation}
with $W$ the Lorentz factor, and $g$ the determinant of the spatial 3-metric.The axisymmetrized
disk used in this work is constructed from $(\tilde \rho, \tilde P, \tilde v^x_T, \tilde v^y_T, \tilde v^z_T)$,
where P is the pressure, and $v^i_T = u^i/u^t$ is the 3-velocity of the fluid with respect to the numerical
grid (``transport'' velocity). We also constructed an axisymmetrized disk from the average temperature
$\tilde T$, and found no significant differences in the subsequent evolution of the system.
We can assess the effect of this axisymmetrization procedure from Figs. 5, 7 and 9 of
Foucart et al. 2015~\cite{foucart2015}, which show the density, velocity, temperature and electron fraction
of the fluid in the equatorial plane of the system $20\,{\rm ms}$ after merger. 
The less symmetric variable is the temperature, which
varies by up to a factor of 2 at constant radius in the equatorial plane. Other fluid variables are
slightly more axisymmetric. 

The dynamical ejecta is reconstructed from the evolution of tracer particles
initialized from an earlier snapshot of the evolution, at $t=4\,{\rm ms}$
after merger. This is early enough that all of the
dynamical ejecta material is still within the computational domain, but late
enough that a Newtonian ballistic and adiabatic approximation for the subsequent
evolution of the particles is acceptable (see e.g.~\cite{roberts2016}, where
smoothed particle hydrodynamics simulations of the tidal ejecta of a BH-NS merger
were performed with and without the inclusion of pressure terms, with little
differences in the evolution of the ejecta). 
Using these assumptions we determine the location of the bound
and unbound components of the dynamical ejecta at $t=20\,{\rm ms}$ 
after merger, and axisymmetrize that mass distribution.
For this component we only consider matter which
is located at a radius $r\gtrsim 9GM_{\rm BH}c^{-2}$ when
the particles are introduced.
This includes all of the dynamical ejecta material which leaves the SpEC
computational grid during the post-merger evolution.
To limit the error introduced by the transition from a general relativistic
simulation to Newtonian trajectories, we determine the velocity of the
tail material by keeping constant the location, direction of motion,
and specific energy of the particles. The latter is estimated using
the approximation $1 + (v/c)^2/2 - GM_{\rm BH}/(rc^2)= -u_t$, where $u^\mu$ is the
4-velocity of the fluid in the general relativistic simulation. 

The disk data is mapped onto a pseudo-Newtonian potential for
long-term evolution (\S\ref{s:flash}) by interpolating the fluid density, pressure,
electron fraction, and velocity, with the remaining
variables obtained with the equation of state self-consistently.
For the dynamical ejecta, the same method is used except that 
the pressure -- which is dynamically unimportant and not extremely reliable 
given the adiabatic and ballistic evolution of particles -- is replaced by 
the entropy of the matter distribution, from which
the temperature is obtained. Regions of very low
particle density are removed from the mapping, with the
resulting composite mass distribution having nearly constant
height-to-radius ratio (Figure~\ref{f:ic}). 

Any regions in the dynamical ejecta
that overlap the disk distribution are removed, and an additional
cut in radius at $r_{\rm cut}=260$~km is made to prevent spurious jumps in 
the density distribution when combining the two components, resulting
in a gap of $\sim 50$~km in radius between disk and dynamical ejecta.
This gap is filled on a timescale of $\sim 10^{-4}$~s once evolution begins, 
and its introduction does not have a significant outcome in the evolution.
For both components, the interpolation is performed after shifting the radial coordinate 
by $0.5r_g$ everywhere, which yields a disk mass that agrees well with that computed in the
GR simulation. A similar shift was used in~\cite{roberts2016}
to provide a better agreement between GR and Newtonian potentials when
mapping data from SpEC simulations of a BH-disk system 
in the damped harmonic gauge to SPH simulations using a
Paczynski-Witta potential. 

The baseline configuration obtained in this way 
has a central BH mass of $8.071M_\odot$ and
spin $\chi = 0.861$, disk mass $0.06M_\odot$, gravitationally
bound dynamical ejecta mass $0.038M_\odot$ (hereafter called \emph{fallback})
and unbound dynamical ejecta mass $0.075M_\odot$ (hereafter called \emph{unbound tail}).
The degree of gravitational binding is computed relative to the pseudo-Newtonian potential.
Passive scalars are assigned to each of these components in order to 
track their evolution and interaction as a function of time (Figure~\ref{f:ic}).
Additional models are evolved which scale the dynamical ejecta mass
or remove it altogether, as described in \S\ref{s:models}.

\subsection{Time-dependent hydrodynamics}
\label{s:flash}

We solve the time-dependent hydrodynamic
equations using FLASH3 \cite{fryxell00,dubey2009} in
axisymmetric spherical polar coordinates $(r,\theta)$. 
The public version of the code has been modified
to include the physics required to model
the long-term evolution of compact object
merger remnants \cite{FM13,FM12,MF14}. Angular 
momentum transport is treated with an
$\alpha$ viscosity prescription \cite{shakura1973},
and neutrinos are included via a leakage scheme
with lightbulb-type self-irradiation, including
only charged-current weak interactions. The equation
of state is that of Timmes et al. \cite{timmes2000}, and gravity
is modeled with the pseudo-Newtonian potential
of Artemova et al. \cite{artemova1996}.

The computational domain extends from $r_{\rm in} = 24.3$~km,
midway between the horizon and the ISCO, until $10^4 r_{\rm in}$,
covering the full range of polar angles. The radial grid is
logarithmic with cell size $\Delta r / r \simeq 3.7\%$, while
the polar grid has 56 cells equispaced in $\cos\theta$, achieving
$\Delta \theta = \Delta r / r \simeq 2^\circ$ on the equator.
We employ standard outflow boundary conditions in radius,
and reflecting in polar angle at the axis.

During the first $1/10$ of an orbit ($\sim 2\times 10^{-4}$~s), the system 
is evolved without source terms to allow sharp edges to smooth 
out and to avoid numerical errors. Subsequently, all source terms are 
included in the evolution. The ambient density is a power-law in 
radius, $\rho_{\rm amb}\propto r^{-2}$, normalized so that it is
nearly $10$ orders of magnitude lower than the torus density peak ($\sim 60$~g~cm$^{-3}$)
at the same location. The density floor is set to be $90\%$ of the initial ambient
density. As time elapses, the floor
is gradually decreased interior to $200$~km until it becomes
constant and equal to $10$~g~cm$^{-3}$.
Source terms are suppressed when the density is less than $10$ times higher than
the floor.

In order to generate homologously expanding ejecta for
radiative transfer calculations, the material flowing out
is sampled at $r_{\rm map} =10^9$~cm throughout the simulation.
Following the approach of \cite{Kasen+15} and \cite{FQSKR2015}, 
the sampled material is injected into a new computational domain
with an inner radius at $r=r_{\rm map}$ and outer radius $r=10^5r_{\rm map}$.
In these new hydrodynamic simulations, material is evolved without energy source terms except for
radioactive heating from $r$-process nucleosynthesis \cite{Korobkin+12}, and with a much 
lower ambient density, $10^{-10}$~g~cm$^{-3}$. By the time
the forward shock reaches $r=3\times 10^{12}$~cm, the velocity
profile is proportional to radius and the kinetic energy
is dominant in most of the material.

\subsection{Nucleosynthesis}
\label{s:skynet}

For each simulation, $10^4$ tracer particles are inserted
in the domain, with random positions that follow the mass distribution (i.e., all
particles represent the same amount of mass).
Thermodynamic trajectories as a function of time are then generated
from each particle by interpolating the corresponding variables
from the grid at each time. In addition to density, temperature, and composition,
we sample neutrino and viscous source terms.

Nucleosynthesis calculations on each trajectory are carried
out with the nuclear reaction network code SkyNet \cite{lippuner2015}.
The network includes 7843 isotopes up to $^{337}$Cn.
Forward strong rates are taken from the JINA REACLIB database
\cite{cyburt2010}, with inverse rates computed assuming detailed balance. 
Spontaneous and neutron-induced fission rates are taken from \cite{frankel1947}, 
\cite{panov2010}, \cite{mamdouh2001}, and \cite{wahl2002}. Most weak rates 
come from \cite{fuller1982}, \cite{oda1994}, and \cite{langanke2000}, or
otherwise from REACLIB. Nuclear masses and partition functions
are taken from REACLIB, which contains experimental data where available or Finite-Range Droplet
Macroscopic model (FRDM, e.g.~\cite{moller1995}) values otherwise.
The network evolves the temperature taking into account source terms due 
to nuclear reactions as computed by SkyNet, as well as viscous heating and 
neutrino interactions (charged current emission/absorption on free nucleons)
from the thermodynamic trajectories.

Computations for each trajectory start from nuclear statistical
equilibrium (NSE). Most particles that initially reside in the disk reach temperatures
$T \geq 10^{10}$~K, and their nucleosynthesis is computed from the last time they
reach this temperature.
Most dynamical ejecta particles have temperatures lower than
$3\times 10^9$~K at all times, making computation of NSE infeasible (and a bad approximation).
We therefore ignore all particles for which the maximum temperature is less than $5\times 10^9$~K, 
and instead set their abundance to the average final yield (at a time $10^{13}$~s) 
of model M14-7-S8 from \cite{roberts2016}.
That calculation uses the same GR merger simulation to produce tracer particles only from dynamical 
ejecta material, and starts the evolution at an earlier time such that particles satisfy $T \geq 5\times 10^9$~K.
For the remaining particles, for which the maximum temperature satisfies $5 \leq (T/10^9\textrm{ K})\leq 10$,
nucleosynthesis starts from the time at which the maximum temperature is reached.
The network calculations performed here extend to a time $1\times 10^9$~s, which is sufficiently
long to obtain relatively stable abundances as a function of mass number $A$ except for the
heaviest nuclei ($A > 210$) undergoing alpha decay and fission. Since the disk contribution
above $A = 130$ is sub-dominant, we consider this to be a reasonable approximation (we ignore
elemental abundance changes as a function of $Z$ that take place at times longer than $10^9$~s).

\subsection{Radiative transfer}
\label{s:sedona}

We calculate synthetic light curves and spectra of the models using the
Monte Carlo radiative transfer  code 
SEDONA \cite{kasen2006}. The calculations use the axisymmetric homologous ejecta profiles
from the hydrodynamical calculations described in \S\ref{s:flash}, 
with a setup similar to \cite{barnes2013} and \cite{Kasen+15}.
The broadband opacity of $r$-process elements is approximated
statistically based on the atomic data of \cite{kasen2013}, and using 
the $Y_e$ from the hydrodynamical simulations as a proxy for the composition.
This is a fair approximation given the
very sensitive dependence of the production of heavy r-process in 
$Y_e$ \cite{Kasen+15,lippuner2015}.  We use  $Y_e < 0.25$ 
as a threshold for the production of lanthanides.

The radiative transfer calculations assume local thermodynamic
equilibrium for the atomic level populations and
ionization state. We use the average time-dependent radioactive heating rate from
\cite{Roberts+11} for all points of the ejecta except those with $Y_e > 0.4$,
which we asssume have zero heating.
We assume 100\% thermalization of the radioactive decay products -- in reality,
non-efficient thermalization will reduce the kilonova luminosity, especially
for times after peak \cite{barnes2016}.
We compute wavelength-dependent light curves for 20 polar
angles, equally-spaced in their cosine, over the range $0$ to $\pi$.
We average the results over three wavelength regions: blue optical
($3500-5000\AA$), red optical ($5000-7000\AA$), and infrared ($1-3\mu$m) 
to construct the specific luminosity in different bands.

\subsection{Models evolved}
\label{s:models}

Table~\ref{t:models} summarizes the properties of the eight models computed in this study. 
The first group includes a simulation of the disk (Fdisk) or dynamical ejecta alone (Fdyn), as well
as a case of disk alone with $Y_e = 0.1$ enforced initially (Fdisk-Y0.1) -- to test the sensitivity
of the wind composition to initial conditions ($Y_e=0.1$ is a common choice
for the initial composition in post-merger simulations starting from idealized quasi-equilibrium tori,
e.g.~\cite{FM13}) -- and a model of disk alone
with no energy source terms (Fdisk-ns), to quantify the contribution of initial transients
to the mass ejection. In all models, a viscosity parameter $\alpha = 0.03$ is used (c.f. \cite{FM13}),
and the initial black hole mass and spin are used in the pseudo-Newtonian potential and kept constant
in time.

The second group of models evolves the combined dynamical ejecta plus disk.
The fiducial case (F0) simulates the system with the physical parameters from the
GR merger simulation, whereas the remaining models (Ft0.1, Ft0.3, and Ft3.0)
scale the dynamical ejecta mass by factors of $1/10$, $1/3$, and $ 3$. While the range of scaling
factors used here is very wide compared to the numerical error in the determination
of the mass of the disk, bound tail, and unbound tail, they are fairly representative
of potential uncertainties in the relative mass of these components when varying the parameters
of the binary (i.e., for black hole masses in the range $M_{\rm BH}\sim [5-10]\,M_\odot$). In particular,
models which reduce the ratio of the mass of the dynamical ejecta to the mass of the disk
are more representative of low-mass BH-NS mergers, for which the disk mass often is an order of magnitude larger
than the mass of the dynamical ejecta (see e.g. Fig.11 of ~\cite{kyutoku2015}). 
\footnote{We note that this statement only applies to the ratio of the disk and dynamical ejecta masses. 
In absolute terms,
and at constant black hole spin, low mass black holes are not necessarily unfavorable to the production
of unbound material. In fact, for low spins, high mass black holes are unable to disrupt the neutron
star and do not produce any ejecta at all~\cite{Foucart2012b}.}
The scaling factor could be even smaller
for lower mass BH-NS mergers ($M_{\rm BH}<5M_\odot$) or NS-NS mergers. 

The evolution of the accretion disk and dynamical ejecta including all source terms
is carried out for $3000$ orbits at the initial density peak, or about $\sim 10$ seconds.
This timescale corresponds to several viscous times, and is determined by demanding
saturation in the mass ejection as measured at the radius at which the disk outflow
is sampled ($r_{\rm map}=10^9$~cm). The second simulation step that uses the
sampled outflow and evolves it into homomology is run until about $100-200$~s. This 
time is chosen so that most of the mass distribution achieves homology while satisfying
the requirement that the swept up mass (in ambient medium) remains lower than
$1\%$ of the disk ejecta mass.

\begin{table}
\caption{Models evolved and summary of results. Columns from left to right show model
name, initial masses of disk, fallback, and unbound tail, fraction of the disk and
fallback ejected in the wind, total ejected fraction $f_{\rm w,tot}$ (eq.~[\ref{eq:fw}]), mass-weighted radial
velocity and electron fraction in the wind (including disk and fallback), mass
ejected with $Y_e > 0.25$, and accreted fractions of disk and fallback.\label{t:models}}
{\scriptsize
\begin{tabular}{lccccccccccc}
\br
Model & $M_{\rm d}$ & $M_{\rm f}$ & $M_{\rm ut}$ & $f_{\rm w,d}$ & $f_{\rm w,f}$ & $f_{\rm w,tot}$ &
        $\bar{v}_{r,{\rm w}}/c$ & $\bar{Y}_{e,{\rm w}}$ & $M_{{\rm w},0.25}$ & $f_{\rm acc,d}$ & $f_{\rm acc,f}$\\
      & \multicolumn{3}{c}{$(10^{-2}M_\odot)$} & $(\%)$ & $(\%)$ & $(\%)$ & 
      $(10^{-2})$ & & $(10^{-3}M_\odot)$ & $(\%)$ & $(\%)$\\
\mr
Fdisk        & 6   &   0  &   0 & 8    & ... & 8    & 3.9 & 0.35 & 5    & 92  & ...\\ 
Fdisk-Y0.1   &     &      &     & 8    & ... & 8    & 3.9 & 0.31 & 4    & 92  & ...\\ 
Fdisk-ns$^a$ &     &      &     & 0.5  & ... & 0.2  & 18  & 0.22 & 0.07 & 8   & ...\\ 
Fdyn         & 0   & 3.8  & 7.5 & ...  & 12  & 12   & 5.0 & 0.07 & 0    & ... & 88 \\ 
\noalign{\bigskip}                                
F0     & 6   & 3.8  & 7.5  & 5    & 32  & 15   & 4.4 & 0.24 & 9 & 95 & 68 \\ 
Ft0.1  &     & 0.38 & 0.75 & 8    & 55  & 11   & 3.5 & 0.30 & 6 & 93 & 42 \\ 
Ft0.3  &     & 1.3  & 2.5  & 7    & 48  & 14   & 3.9 & 0.26 & 8 & 94 & 52 \\ 
Ft3.0  &     & 11   & 23   & 5    & 20  & 15   & 4.6 & 0.20 & 8 & 96 & 81 \\ 
\br
\end{tabular}\\
}
$^a$ Model Fdisk-ns is evolved without momentum or energy source terms.
\end{table}

\section{Dynamics of disk wind and dynamical ejecta}

\subsection{Overview of baseline model}

The initial condition for the baseline model F0 is shown in 
Figure~\ref{f:ic}. Passive scalars track the evolution of 
the disk material as well as the gravitationally bound and unbound parts
of the dynamical ejecta (\emph{fallback} and \emph{unbound tail}). For clarity, we start by discussing the 
behavior of models that include only the disk or the dynamical ejecta, and then 
proceed to describe their combined evolution. 

At $t=0$, the electron fraction in the disk and dynamical ejecta is very 
different. With the exception of very low density regions near its inner
edge, most of the dynamical ejecta has $Y_e \simeq 0.05$. In contrast,
the bulk of the disk has $Y_e \simeq 0.15-0.2$, increasing away from
the disk midplane. The initial electron fraction of the disk is somewhat
higher than that employed in long-term disk simulations that start 
from equilibrium tori as initial conditions (e.g., \cite{FM13}).

\begin{figure*}
\includegraphics*[width=0.5\textwidth]{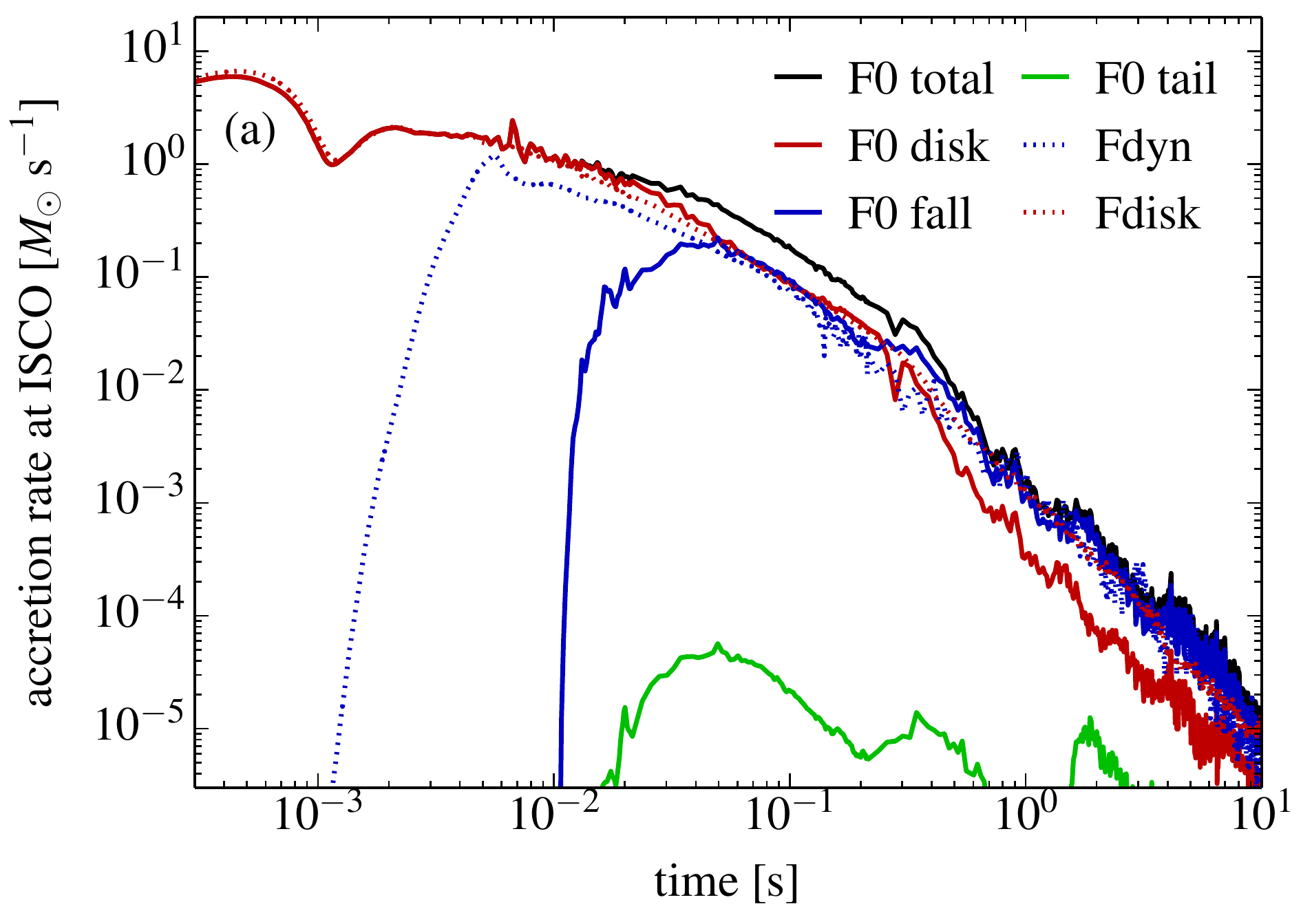}
\includegraphics*[width=0.5\textwidth]{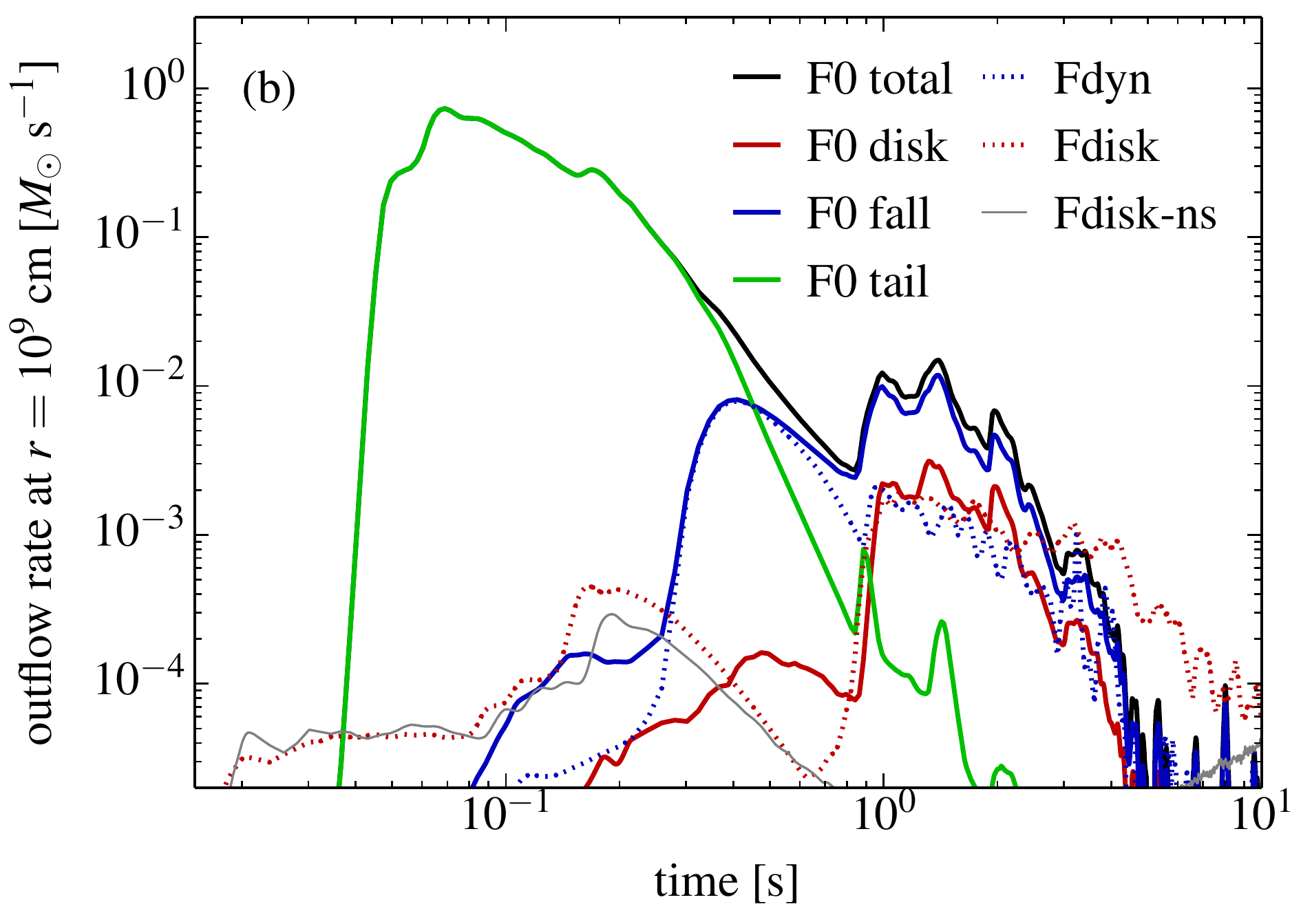}
\caption{\emph{Left:} Mass accretion rate at the ISCO for the different
components of the baseline model F0 shown as solid lines: disk (red), 
\emph{fallback} (blue), and \emph{unbound tail} (green). 
For comparison, dotted lines show accretion of disk material for the model without
dynamical ejecta (Fdisk, red) and of \emph{fallback} material for the model without disk (Fdyn, blue).
\emph{Right:} Same as the left panel but now showing total mass ejection at
$r=10^9$~cm. The disk-only model without source terms (Fdisk-ns) is shown with a thin grey line.}
\label{f:mass_fluxes_tidal}
\end{figure*}

In the absence of dynamical ejecta (model Fdisk), the disk follows the usual
stages of neutrino-cooled disks (e.g., \cite{metzger2008steady}). Two time scales
govern the evolution of the system: the \emph{orbital time}
\begin{equation}
t_{\rm orb} \simeq 2\left(\frac{8M_\odot}{M}\right)^{1/2}\left(\frac{r}{55\textrm{ km}} \right)^{3/2}~\textrm{ms},
\end{equation}
where $M$ is the black hole mass, and the \emph{viscous time}
\begin{equation}
t_{\rm visc} \simeq 200\left(\frac{0.03}{\alpha}\right)\left(\frac{0.25}{H/R} \right)^2
\left(\frac{8M_\odot}{M}\right)^{1/2}\left(\frac{r}{55\textrm{ km}} \right)^{3/2}~\textrm{ ms},
\end{equation}
where $H/R$ is the height-to-radius ratio of the disk. Initially,
neutrino cooling is important, balancing viscous heating. As the disk evolves
and spreads, neutrino emission gradually decreases. At the same time, nuclear
recombination adds energy in the outer parts of the disk. By $t=t_{\rm visc}\simeq 200$~ms, 
the neutrino luminosity has decreased by
two orders of magnitude from its initial value, and the disk has reached an
\emph{advective} state, leading to vigorous convection and mass ejection. 
As shown in Table~\ref{t:models} however, most of the initial disk mass
is accreted onto the BH (Figure~\ref{f:mass_fluxes_tidal}a), with only $\sim 8\%$ 
ejected as a wind. The bulk of mass ejection reaches $10^9$~cm at times $\gtrsim 1$~s and 
later (Figure~\ref{f:mass_fluxes_tidal}b).
A small amount of mass due to the non-equilibrium initial condition is ejected on the 
thermal time ($t\lesssim 30$~ms), this
outflow is also present in a comparison model evolved without any source terms 
other than gravity (Fdisk-ns).

In the absence of an accretion disk (model Fdyn), most of the
\emph{fallback} matter moves quickly towards the BH, 
while the \emph{unbound tail} moves outward. Both components also expand in
polar angle, due to small pressure gradients neglected in the ballistic evolution of the
dynamical ejecta material. Initially, the \emph{fallback} 
material has nearly constant specific angular momentum, in excess
of the value at the innermost stable circular orbit (ISCO) by a factor of 
$\sim 1.3$. Therefore, the part of this material that falls
towards the BH forms an accretion disk, 
and the resulting accretion rate at the ISCO evolves
in a way similar to the original accretion disk after \emph{fallback} material has
reached the ISCO (Figure~\ref{f:mass_fluxes_tidal}a). 
Meanwhile, the \emph{unbound tail} and part of the \emph{fallback} matter 
move continuously outward. Over the course of a few orbits, the outward-moving dynamical
ejecta establishes a power-law density profile in radius that transitions smoothly
between bound and unbound components. A very small amount of \emph{unbound 
tail} material is mixed within the \emph{fallback} at small radius,
and moves towards the BH due to lack of angular momentum. As shown in Table~\ref{t:models},
12\% of the initial \emph{fallback} mass reaches a radius of $10^9$~cm (9\% gravitationally
unbound), while the rest is accreted. A fraction 99.6\% of the initial \emph{unbound tail} mass leaves the system.

\begin{figure*}
\includegraphics*[width=0.95\textwidth]{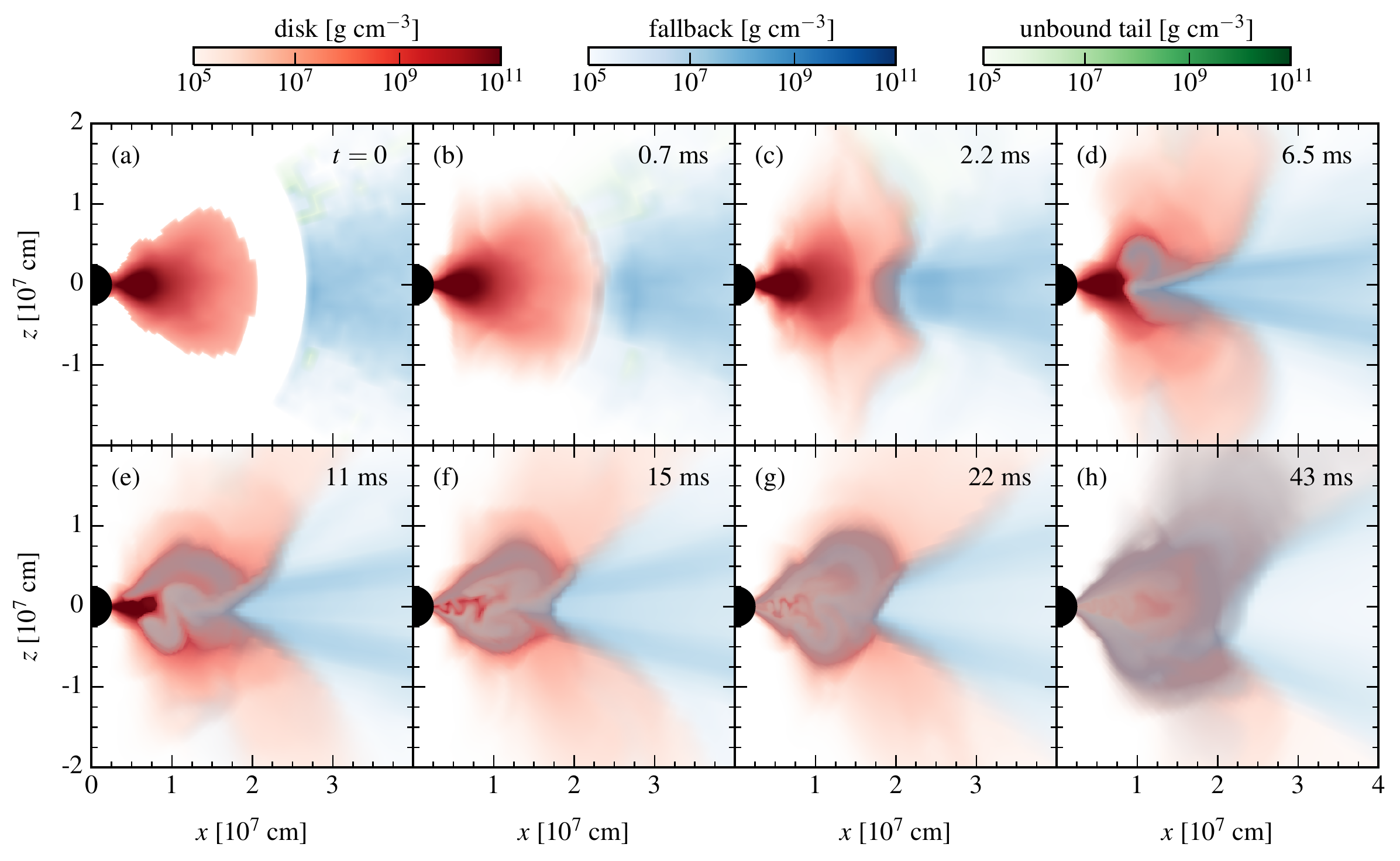}
\caption{Snapshots of the early evolution of the baseline model F0, showing how the \emph{fallback}
material mixes into the disk. The color maps show partial densities $\rho X_i$, with $X_i$
the mass fraction of disk, \emph{fallback}, or \emph{unbound tail} material, with an opacity that varies
linearly from $0$ to $1$ alongside the color table. Note that the density range for the color
scale differs from Figure~\ref{f:ic}; here it is the same for all components. The orbital
time at the location of the initial disk density peak is $2.2$~ms.}
\label{f:mixing}
\end{figure*}

The early evolution of the combined disk and dynamical ejecta (model F0) is illustrated
in Figure~\ref{f:mixing}. The gap that initially separates the outer edge of the 
disk and the inner edge of the bound tail is filled on a time $t < t_{\rm orb}$ by expansion of the disk and
inward motion of the bound tail. \emph{Fallback} matter continues to penetrate the disk
on the equatorial plane at later times, with mixing of the two components occurring
on a timescale of $\sim 10$ orbits at the disk density peak. For comparison, model Ft0.1
(dynamical ejecta density scaled down by a factor of $10$ relative to the
baseline model F0) does not experience this violent mixing. Instead, the \emph{fallback}
matter mixes only around the periphery of the disk and does not penetrate regions
of higher density. This behavior confirms that the initial gap between disk and dynamical
ejecta does not influence the dynamics significantly. Instead, the relevant
parameter is the ratio of dynamical ejecta mass to disk mass (which is related to their
densities).

A snapshot of the longer-term expansion of the system -- compared to the
cases in which the disk and dynamical ejecta evolve independently -- is
shown in Figure~\ref{f:partial_dens}a. The bulk of the \emph{unbound tail} material 
moves outward identically regardless of whether the disk is present. 
By $t = t_{\rm visc}\simeq 200$~ms, the \emph{fallback} material has fully mixed into
the original accretion disk, as shown by the similar shape of the density profiles of the
two components inside $r\sim 200$~km. The original disk material is 
compressed by the infalling matter, maintaining a 
sharp density drop at its initial outer edge ($200$~km), and a higher density peak relative 
to the case without dynamical ejecta at the same evolution time. Nevertheless, 
the very outer edge of the disk has still made its way into the dynamical ejecta, reaching
a similar radius ($\sim 2000$~km) as model Fdisk, at the same
evolutionary time. The disk outflow of model F0 also contains material initially labeled as
\emph{fallback} in higher abundance than disk material outside $r\sim 200$~km.
These results indicate that (a) with the exception of the outermost regions of the dynamical ejecta,
the evolution of which is unaffected by the disk,
the \emph{fallback} and disk components mix efficiently (for model F0), becoming effectively a single
disk, and that (b) the wind ejection mechanism is the same as in the case
without dynamical ejecta. Changes in 
the wind properties must therefore be a consequence of the different
initial binding energies and masses of the disk and bound dynamical ejecta components.

\begin{figure*}
\includegraphics*[width=0.95\textwidth]{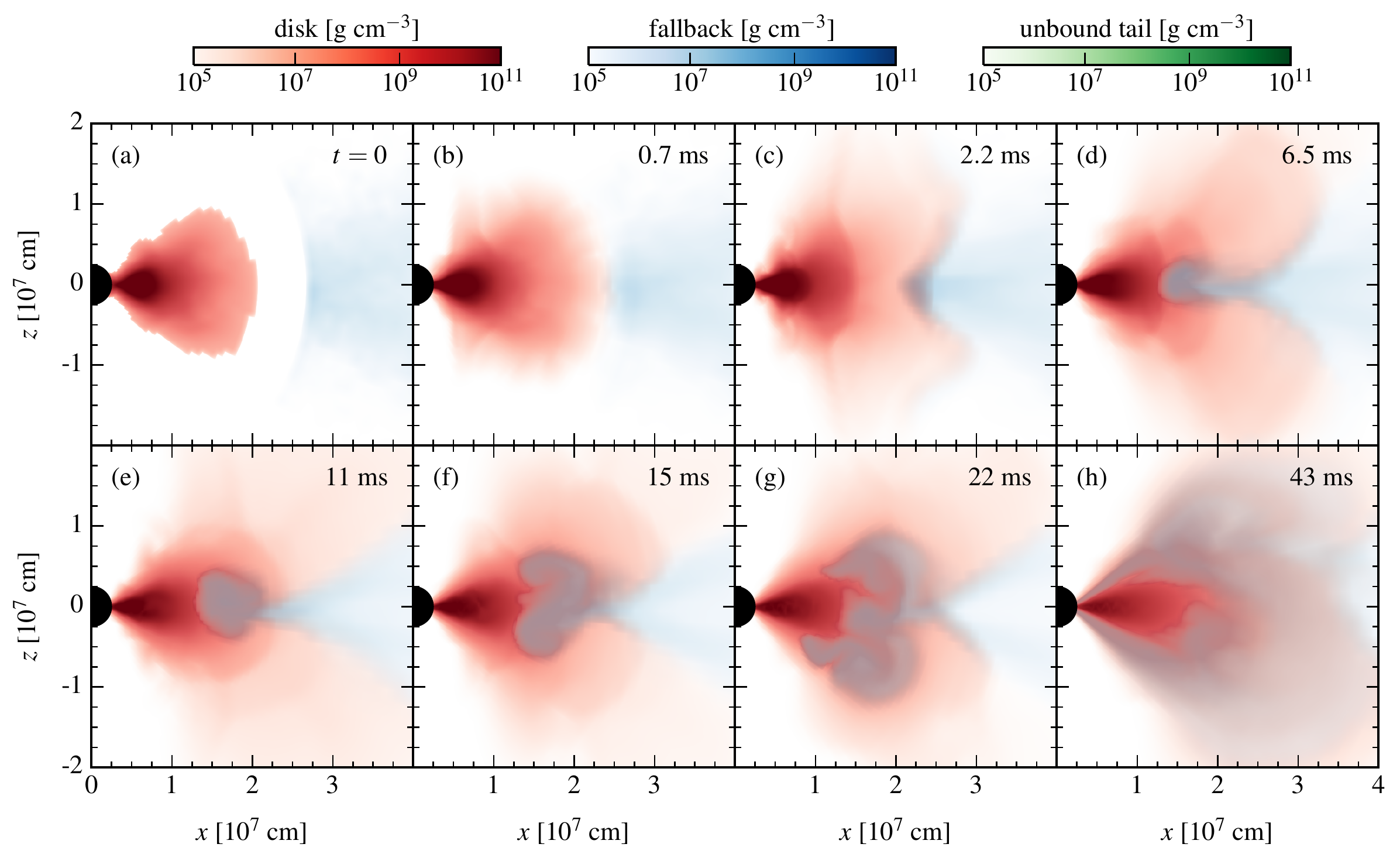}
\caption{Same as Figure~\ref{f:mixing}, but for model Ft0.1. The \emph{fallback} matter
has a lower density than in model F0, and only mixes in the periphery of the disk.}
\label{f:mixing_x01}
\end{figure*}

\begin{figure*}
\includegraphics*[height=1.85in]{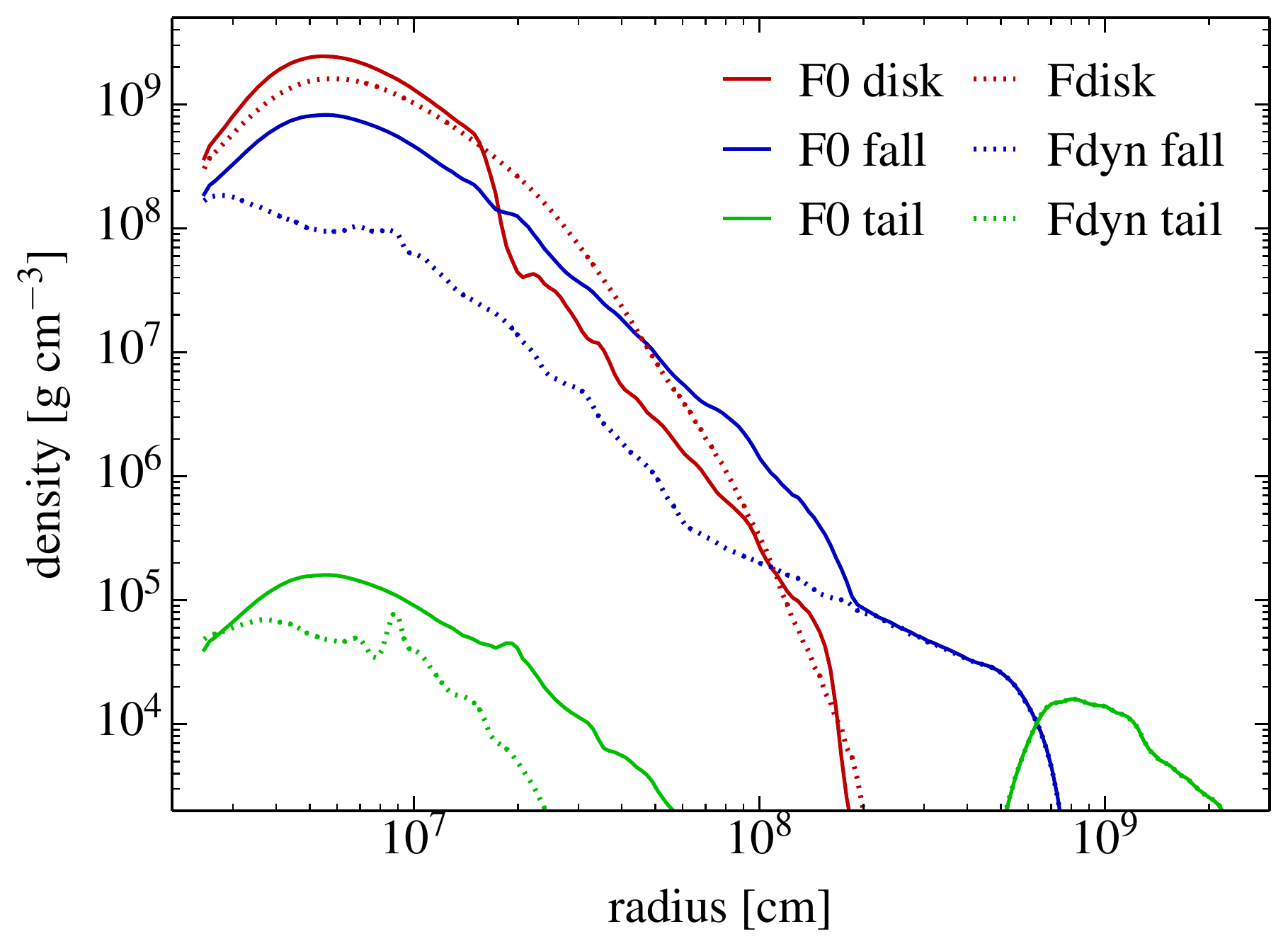}
\includegraphics*[height=1.85in]{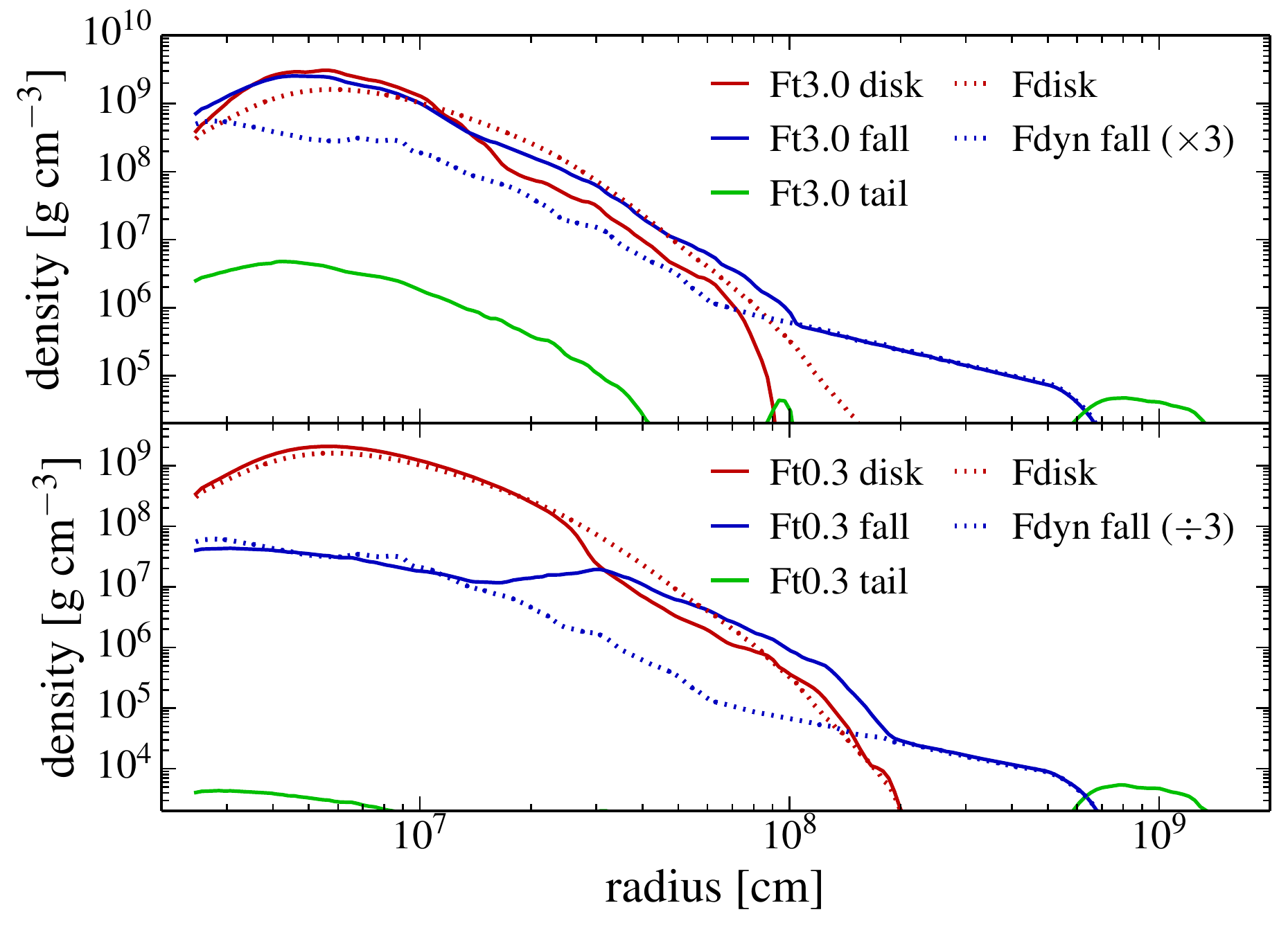}
\caption{\emph{Left:} Angle-averaged density profiles for the different components of 
the disk-dynamical ejecta remnant system at $t=216$~ms: disk (red), \emph{fallback} (blue), 
and \emph{unbound tail} (green). For comparison, dotted lines
show profiles for the models with disk alone (Fdisk, red) and dynamical ejecta 
alone (Fdyn, blue and green) at the same time.
\emph{Right:} Same as left panel but for models with scaled dynamical ejecta at 
$t=216$~ms: Ft3.0 (top) and Ft0.3 (bottom). Models
with disk and dynamical ejecta alone (Fdisk and Fdyn, respectively) are shown with dotted lines.
The curves from model Fdyn are scaled by the same factor as models Ft0.3 and Ft3.0 (as labeled).}
\label{f:partial_dens}
\end{figure*}

\subsection{Effect of relative mass of dynamical ejecta and disk}

\begin{figure*}
\includegraphics*[width=0.5\textwidth]{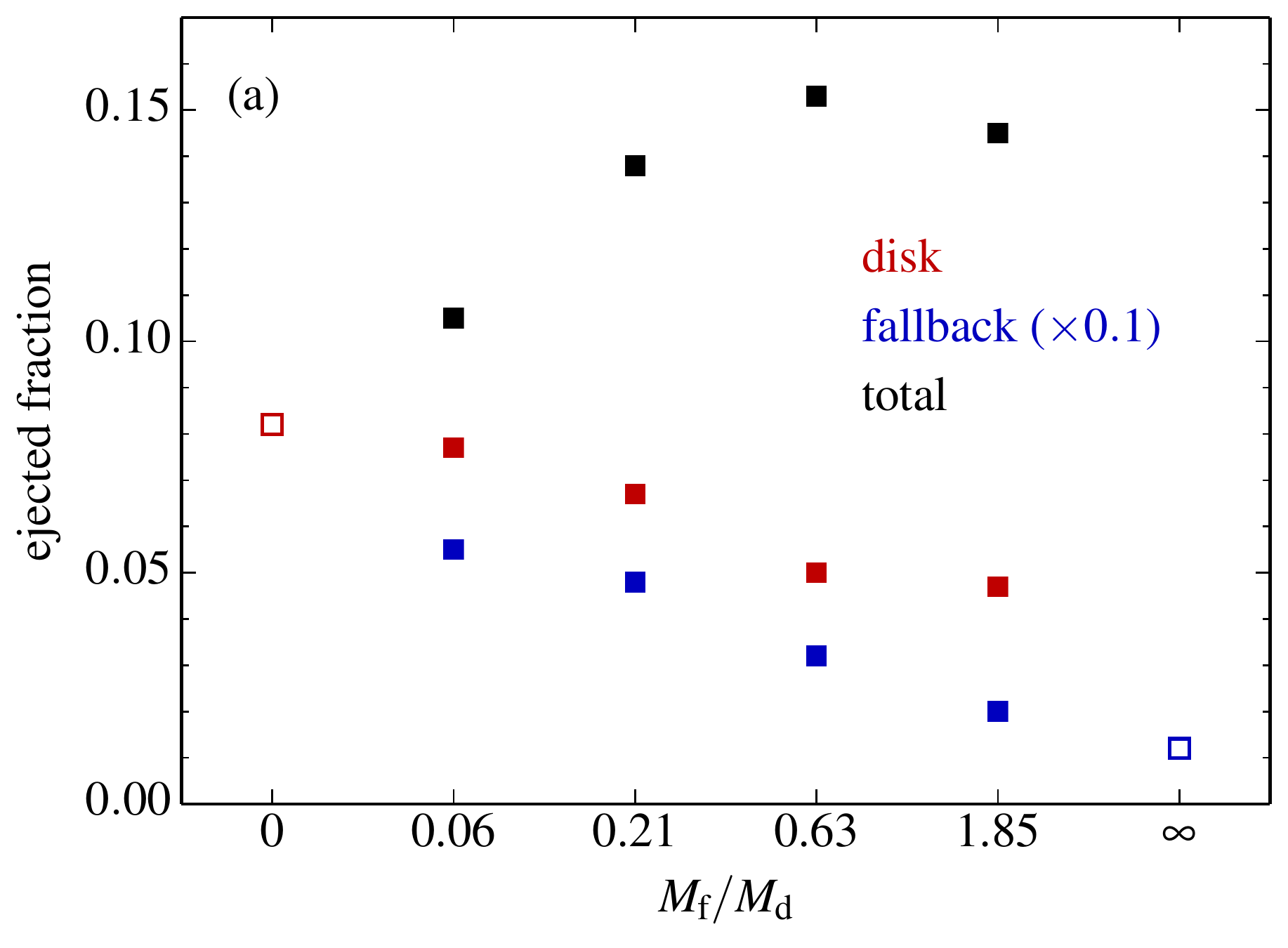}
\includegraphics*[width=0.5\textwidth]{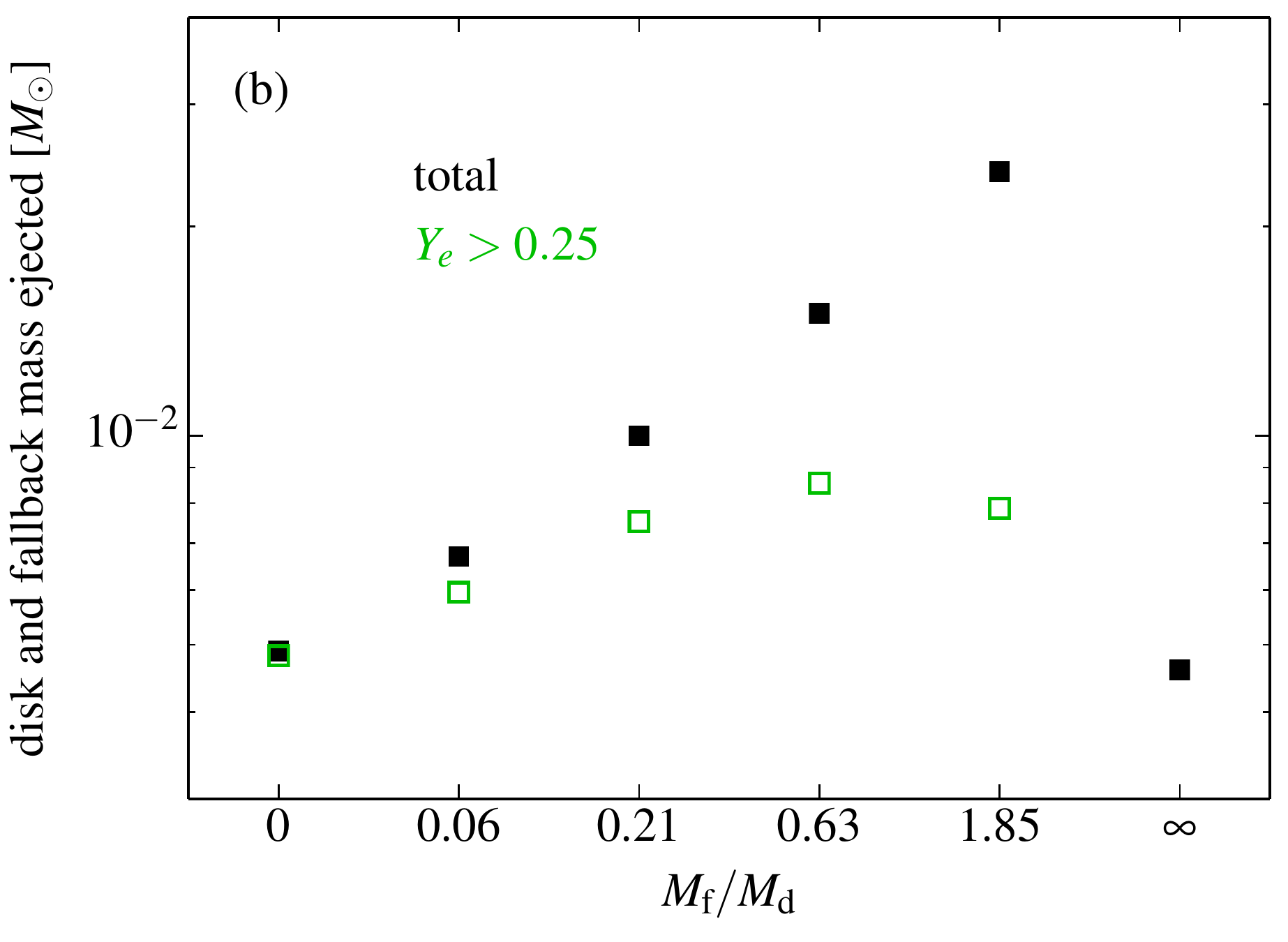}
\caption{\emph{Left:} Fraction of the initial mass in disk (red), \emph{fallback} (blue), 
and total disk plus \emph{fallback}
material (black, eq.~[\ref{eq:fw}]) that is ejected as a wind at $r=10^9$~cm, 
as a function of the ratio of initial \emph{fallback} to
disk masses, for constant initial disk mass (c.f. Table~\ref{t:models}). Note
that the fraction of \emph{fallback} matter ejected is scaled down by a factor
of $10$ to fit in the plot. \emph{Right:} Total mass (black)
and mass with $Y_e >0.25$ (green) ejected in the disk wind as a function of the ratio of initial \emph{fallback} to
disk masses. The mass with $Y_e > 0.25$ is a proxy for the amount of Lanthanide- and Actinide-free mass.
See Table~\ref{t:models} for numerical values.}
\label{f:ejected_fractions}
\end{figure*}

The overall trends in mass ejection as a function of initial
\emph{fallback} and disk masses are summarized in Table~\ref{t:models} and Figure~\ref{f:ejected_fractions}a.
Ejection is quantified by the ratio of total ejected mass ($M_{w,i}$) at $r=10^9$~cm 
to the initial mass ($M_i$) in disk or \emph{fallback components}, $f_{w,i} = M_{w,i}/M_i$, with $i=\{d,f\}$ 
for disk or fallback, respectively. We also define a total ejected fraction of disk plus fallback material,
\begin{equation}
\label{eq:fw}
f_{w,tot} = \frac{f_{w,d}M_d + f_{w,f}M_f}{M_d + M_f}.
\end{equation}
Note that in computing the ejected mass, we include all material that reaches $r=10^9$~cm, 
irrespective of whether it is gravitationally unbound or not. Equation~\ref{eq:fw} is
equivalent to adding up the mass of disk and \emph{fallback} material ejected and dividing
the result by the sum of initial disk and \emph{fallback} masses.

The fractions of ejected disk and fallback material both decrease for increasing $M_f/M_d$, 
while the total fraction increases to a peak of $f_{w,tot}\sim 15\%$ around $M_f \sim M_d$.
This behavior arises from the large fraction of fallback material that is turned around
whenever a disk is present ($>20\%$), and the relatively weak dependence of the 
ejected disk fraction on dynamical ejecta mass. 

To clarify the dynamics underlying these trends, we show in Figure~\ref{f:partial_dens}b
angle-averaged profiles of disk and \emph{fallback} densities for models
with small and large $M_f/M_d$ ratios (Ft0.3 and Ft3.0, respectively). Also 
shown for comparison in each panel is the \emph{fallback} profile in the model with no disk
(Fdyn), scaled to the appropriate dynamical ejecta normalization, and the disk profile in the
model with no tail (Fdisk). When the \emph{fallback}
mass is small, bound tail material causes only minor modifications to the
radial disk profile, with most of the change occurring in the outer layers.
In contrast, for a large initial $M_f/M_d$ (model Ft3.0), 
\emph{fallback} material mixes very effectively into the disk, compressing it
significantly relative to the case without dynamical ejecta.

The presence of the disk stops \emph{fallback} from immediately reaching the
BH and delays the onset of accretion of this material (as in e.g.
Figure~\ref{f:mass_fluxes_tidal}b). This partially accounts for the fact that the ratio 
of ejected to accreted matter in \emph{fallback} material increases as $M_f/M_d$
decreases. Additionally, the (negative) net specific energy of the fluid is initially an order
of magnitude larger at the disk density peak than within the bound
dynamical ejecta material. The fallback component is thus
less gravitationally bound than the disk component, and is relatively easier to eject.

\subsection{Ejecta composition and kinematics}

Table~\ref{t:models} shows the mass-flux-weighted radial velocity and 
electron fraction of the disk and \emph{fallback} material ejected at $r=10^9$~cm for all
models evolved. For increasing $M_f/M_d$, the outflow
is faster and more neutron rich. This trend derives from the fact that
the \emph{fallback} is initially very neutron rich and less gravitationally bound (per unit
mass) than the disk material. When both components mix, a larger mass-weighted outflow velocity
is obtained for fixed energy input per unit mass. Also, material near the transition
between bound and unbound dynamical ejecta moves outward and contributes to increase the 
total outflow momentum. 

\begin{figure*}
\includegraphics*[width=0.5\textwidth]{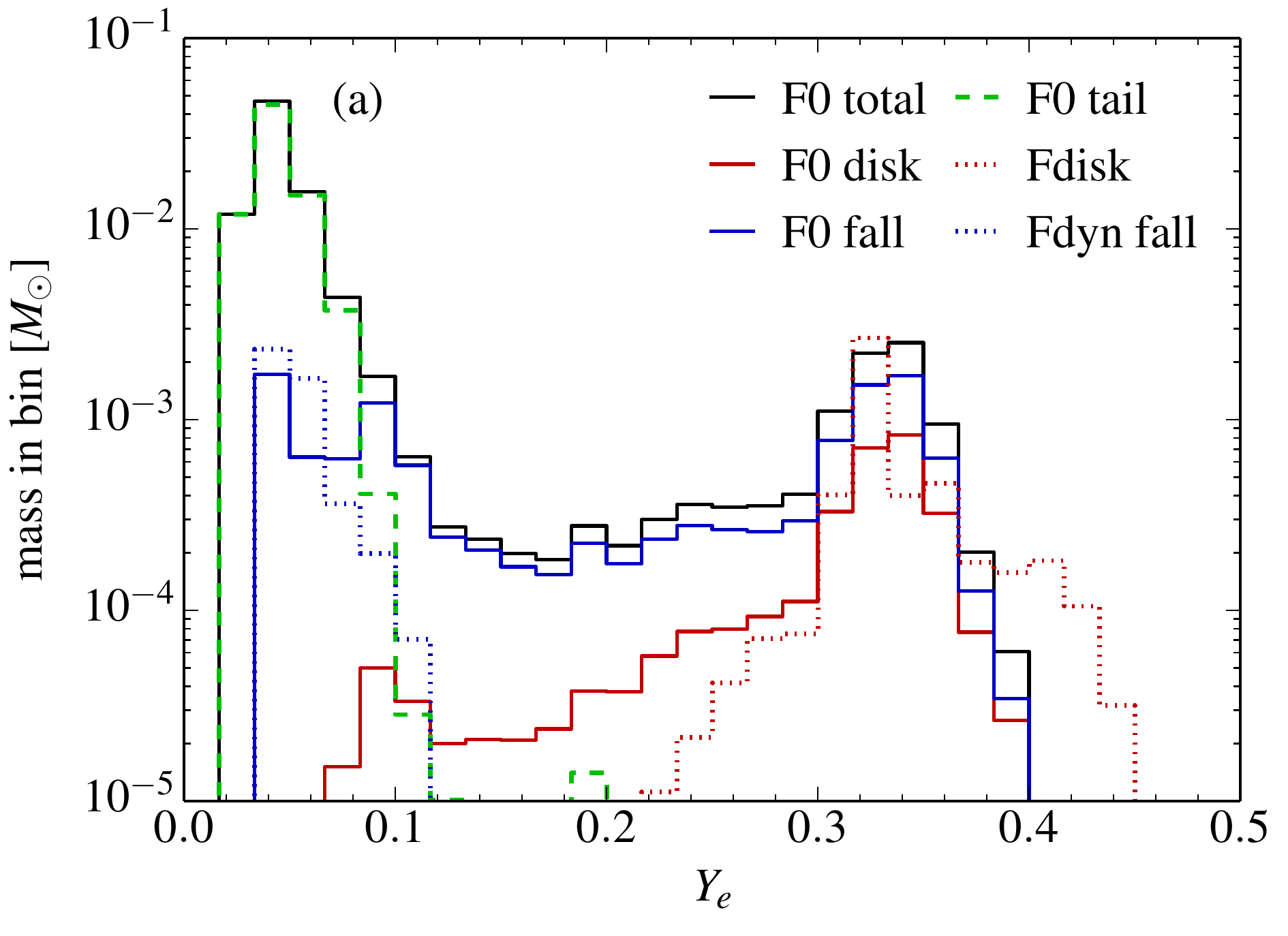}
\includegraphics*[width=0.5\textwidth]{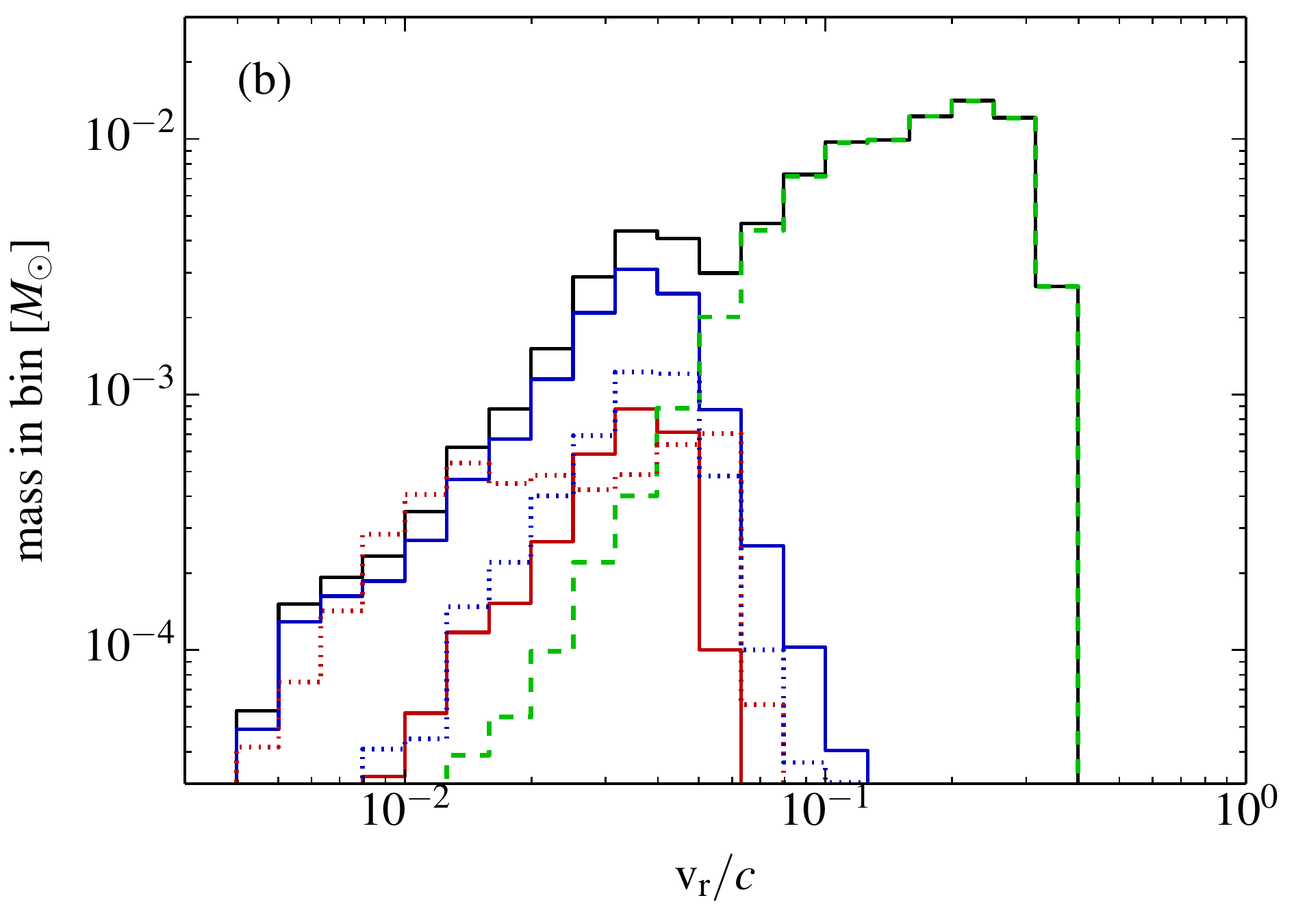}
\caption{Mass histograms of material ejected at $r=10^9$~cm
as a function of electron fraction (left) and radial velocity (right). 
Solid lines and dashed lines correspond to the baseline model F0, while dotted
lines show the disk-only case (Fdisk, red) and dynamical ejecta-only model (Fdyn, blue), for
comparison.}
\label{f:histograms_tidal}
\end{figure*}

For a more detailed breakdown of the outflow properties, Figure~\ref{f:histograms_tidal}
shows mass histograms of electron fraction and radial velocity for the
baseline model F0, and the models without dynamical ejecta (Fdisk) and without
disk (Fdyn), for comparison. The bulk of the \emph{unbound tail} satisfies 
$Y_e \lesssim 0.1$ and is faster than the rest of the outflow, thus evolving independently. 
In the absence of a disk, the \emph{fallback} component has a similar
electron fraction distribution as the \emph{unbound tail}, and its
velocity distribution is a continuation of the unbound tail distribution to lower
velocities. When the disk is present, the \emph{fallback} material acquires
a broad $Y_e$ distribution that stretches from the initial unbound tail region to the 
disk $Y_e$ distribution. Material that mixes deep into the disk is subject to
weak interactions, acquiring a similar composition as disk material, while
matter that remains outermost in radius preserves the original dynamical ejecta
composition, with a smooth transition arising from the gradual decrease in strength
of weak interactions with increasing radius (temperature). The disk $Y_e$ distribution
acquires a low-$Y_e$ tail stretching to $Y_e \sim 0.06$, and the
high-$Y_e$ cutoff is decreased from $0.45$ to $0.4$. 
The velocity distribution of the \emph{fallback} material in the baseline model F0
is essentially a scaled-up version of the distribution of the disk component
in the same model, again pointing to mixing. The high velocity end is dominated
by turned-around \emph{fallback matter}, which replaces matter at the outer edge of the disk
as the fastest and earliest ejecta in the polar direction (relative to model Fdisk).
The disk component of model F0 loses the low-end of its velocity 
distribution relative to model Fdisk. The low-velocity end is now dominated
by outward-moving \emph{fallback} matter far from the bound-unbound transition.

Figure~\ref{f:ejected_fractions}b shows ejected mass with $Y_e > 0.25$, which is 
indicative of material that cannot form Lanthanides and Actinides and therefore has 
an optical opacity similar to iron-group elements, thus contributing to an
optical kilonova signal (e.g. \cite{Kasen+15}). Even though the outflow becomes on average less neutron rich 
with increasing $M_f/M_d$, there is more Lanthanide-free mass ejected until $M_f \sim M_d$,
due to the increasing total mass of the outflows. 

\begin{figure*}
\centering
\includegraphics*[width=0.49\textwidth]{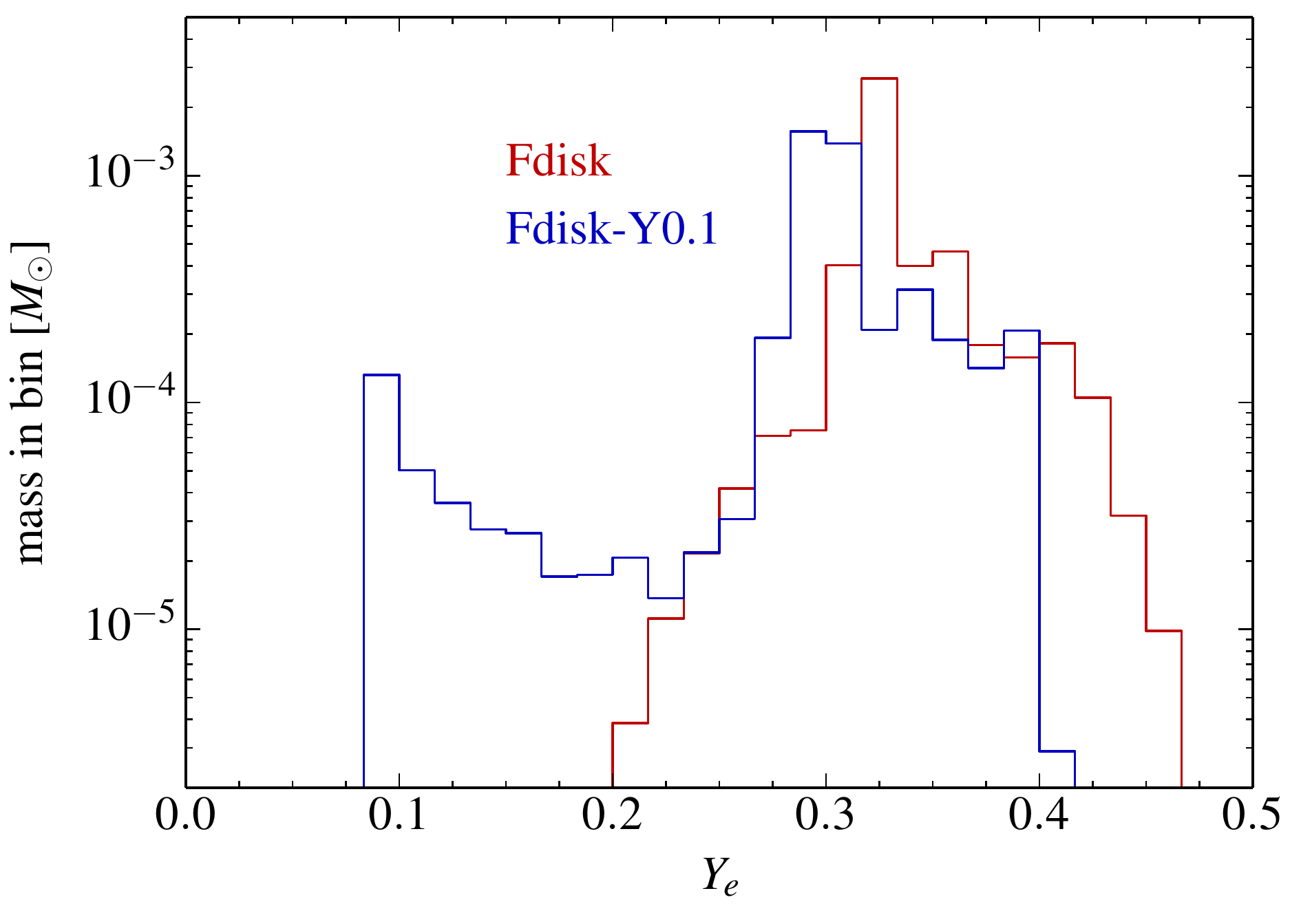}
\includegraphics*[width=0.49\textwidth]{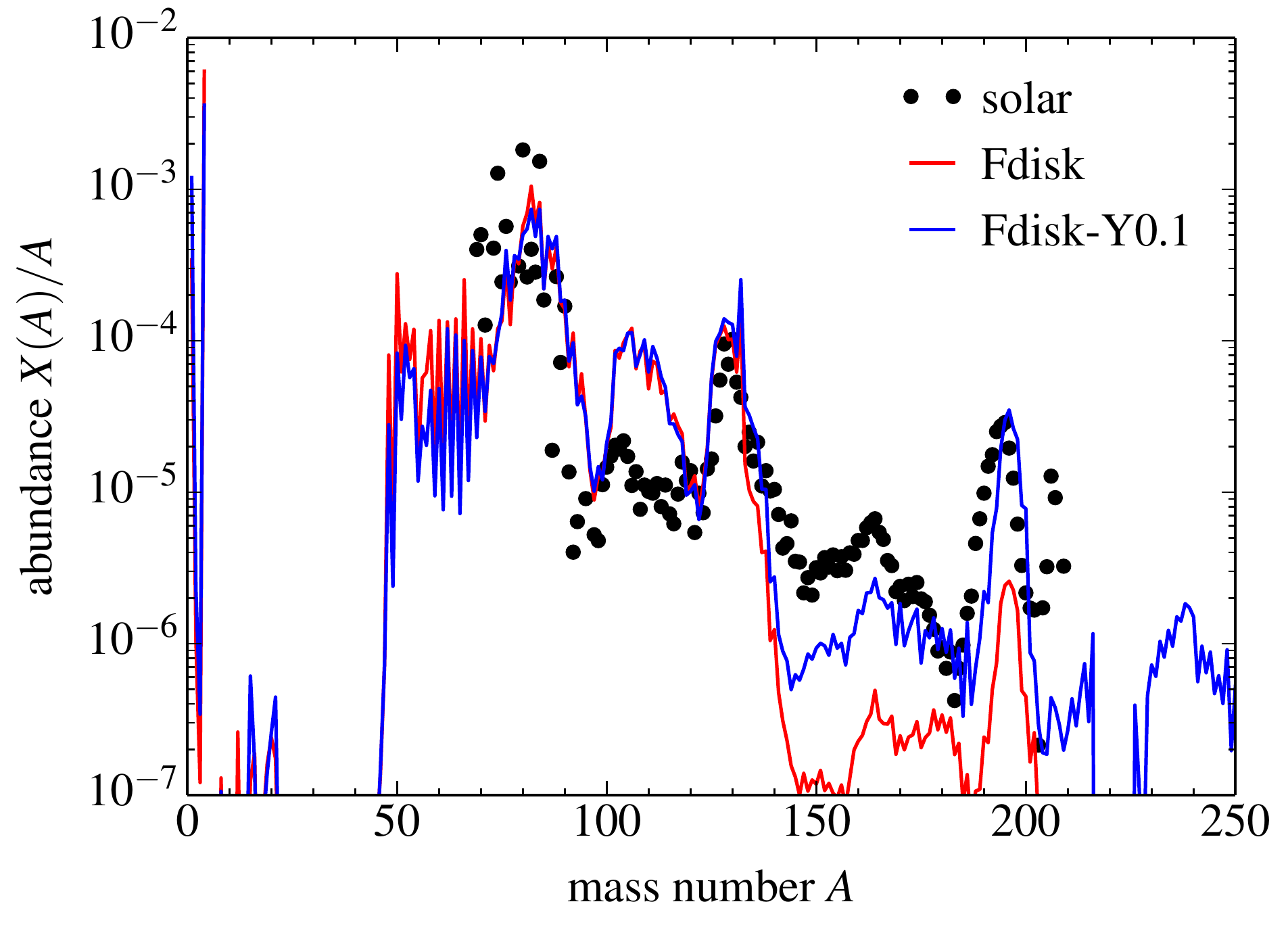}
\centering
\caption{\emph{Left:} Mass histogram as a function of electron fraction for outflow material
measured at $r=10^9$~cm in disk-only models that use an initial $Y_e$ from the merger
simulation (Fdisk) or that impose $Y_e=0.1$ initially (Fdisk-Y0.1). \emph{Right:} Nucleosynthesis
abundances obtained with thermodynamic trajectories from models Fdisk and Fdisk-Y0.1, normalized
so that the mass fractions $X(A)$ integrate to unity. The Solar System
abundances from \cite{goriely1999} are scaled to model Fdisk at $A=130$.}
\label{f:histogram_ye}
\end{figure*}

Finally, we explore the effect of the initial conditions on the 
composition of the disk outflow. Figure~\ref{f:histogram_ye} compares
the $Y_e$ distribution of two models that evolve the disk alone: one
using the electron fraction from the merger simulation (Fdisk), and
another one in which $Y_e = 0.1$ is initially imposed throughout the disk
(Fdisk-Y0.1). The latter prescription is adopted in many long-term simulations of
disk outflows that begin with equilibrium initial conditions 
(e.g., \cite{FM13,just2014,FKMQ14}). Figure~\ref{f:histogram_ye} shows that the
bulk of the outflow has similar average electron fraction properties in both cases, 
with mass-flux weighted averages of $0.35$ and $0.31$ for Fdisk and Fdisk-Y0.1, respectively. 
Aside from the slightly lower mean value for model Fdisk-Y0.1, its distribution has an 
extended low-$Y_e$ tail and a sharper high-$Y_e$ cutoff relative to model Fdisk. 

The insensitivity to initial value of $Y_e$ can be contrasted with the results of simulations
starting from less compact accretion disks, as is obtained in NS-NS mergers.
For example, \cite{FM13} simulated
disks with an initial density peak at $r_{\rm peak}\sim (10-15) GM_{\rm BH}/c^2$,
as opposed to $r_{\rm peak}\sim 5 GM_{\rm BH}/c^2$ here, with the
same code and physical model, finding an average electron fraction
$Y_e\sim 0.15-0.2$. This is significantly different than the results
obtained here. The compactness of the disk is determined primarily by the mass of the black
hole. General relativistic simulations yield density peaks of the post-merger disk at 
$r_{\rm peak}\sim 50\,{\rm km}$ (once a mostly axisymmetric disk forms at times 
$10-20\,{\rm ms}$ after merger), without much dependence on the mass of the black 
hole~\cite{foucart2011,foucart2012}. Accordingly, more massive black holes have more
compact disks.

We conclude that the bulk of the disk outflow acquires its composition from the 
compactness of the disk, which regulates the temperature and therefore
the weak interaction timescale and beta equilibrium $Y_e$. Memory of the initial conditions
imparts modifications at the $\sim 10\%$ level in the electron fraction distribution.

\section{Nucleosynthesis}

\begin{figure*}
\centering
\includegraphics*[width=0.49\textwidth]{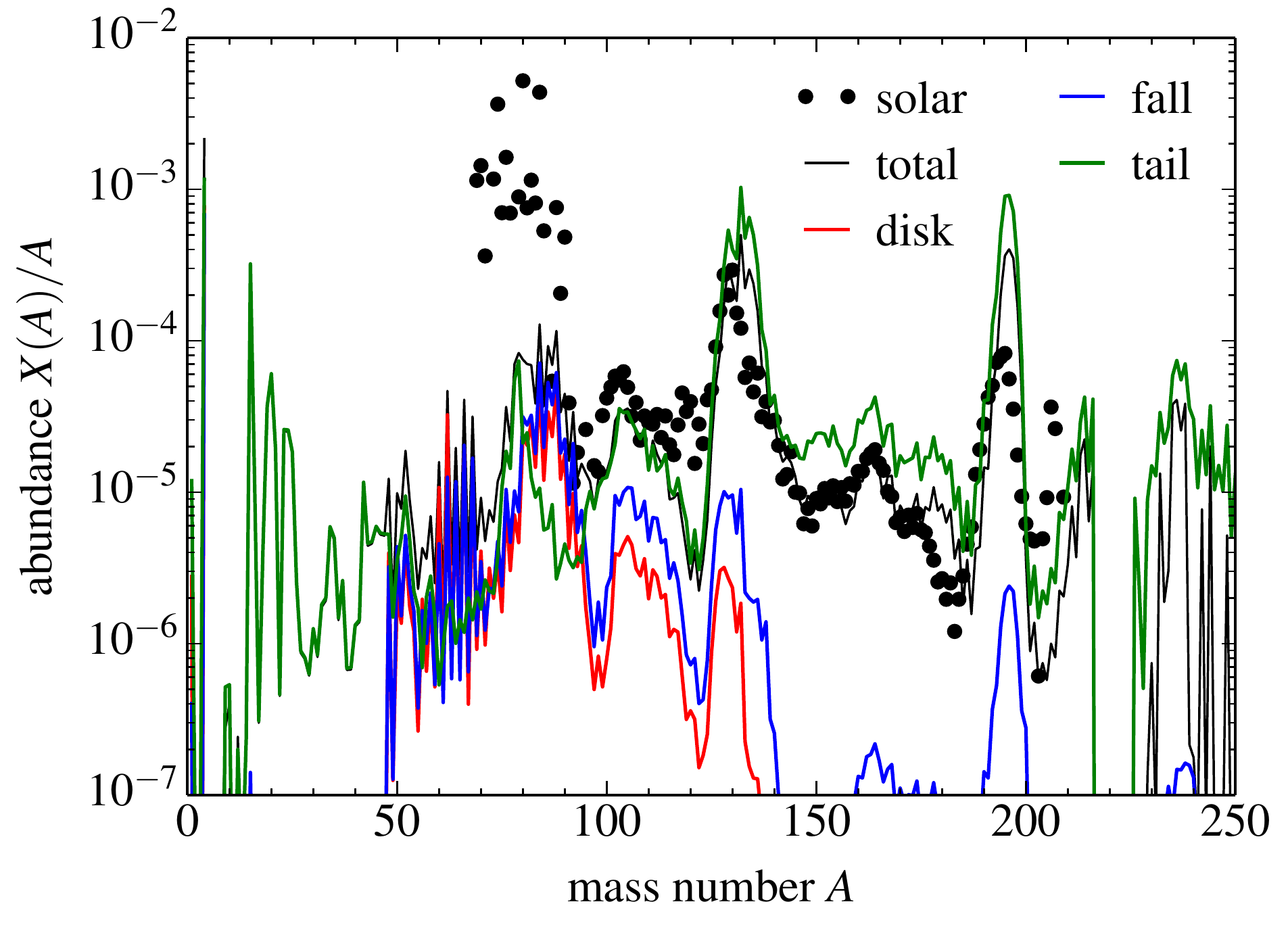}
\includegraphics*[width=0.49\textwidth]{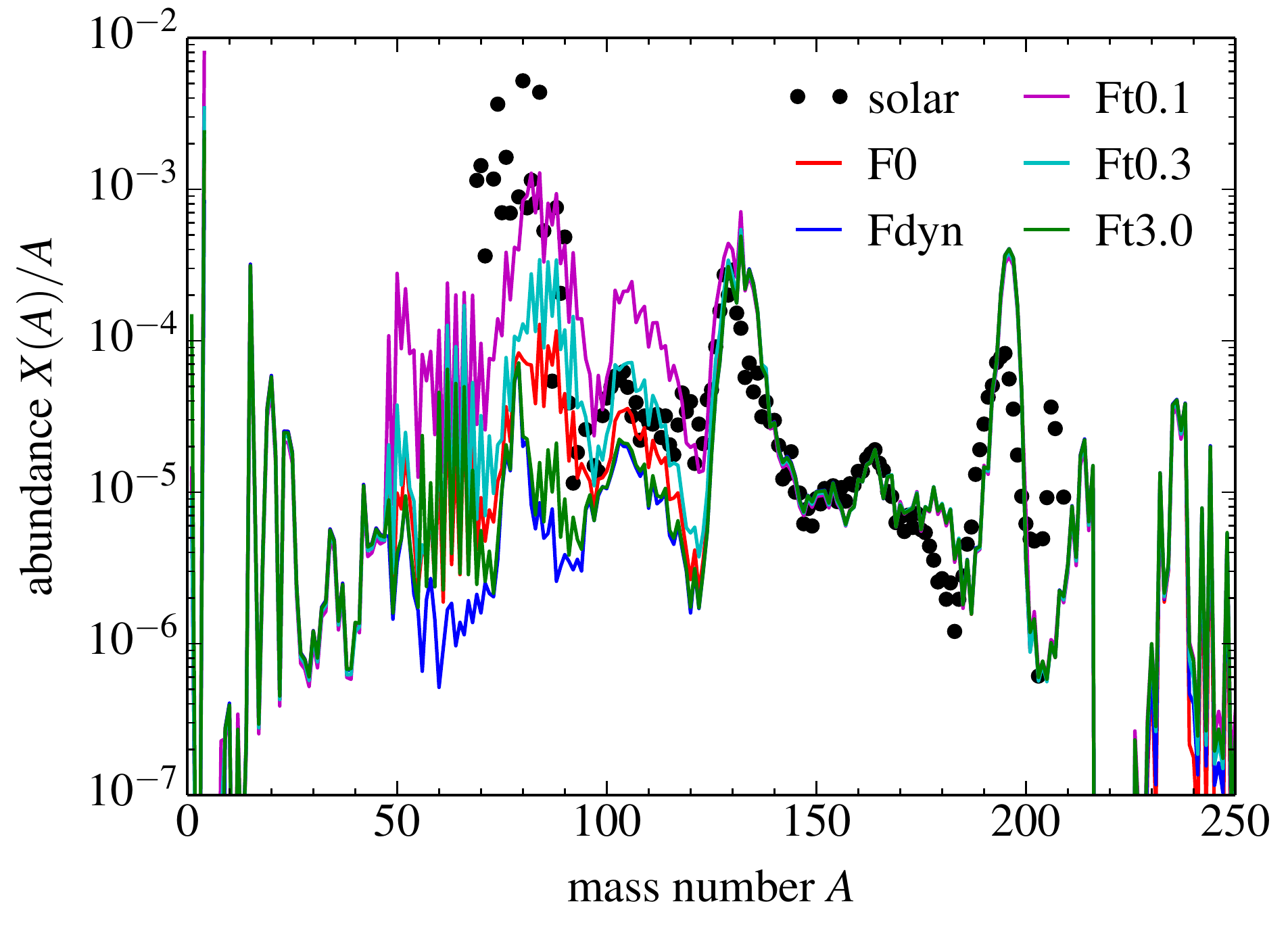}
\centering
\caption{\emph{Left:} Abundances for model F0 broken down by components and weighted
by their relative masses. \emph{Right:} Combined abundance distributions
for models with scaled dynamical ejecta. Model abundance curves and Solar System values 
are normalized to the rare-Earth peak ($A=164$) of model F0.}
\label{f:abundances_combined}
\end{figure*}

Figure~\ref{f:abundances_combined}a shows final abundances for model F0 as a whole and
separated  by ejecta component. The unbound tail produces
heavy $r$-process elements, with a deficit below $A \sim 90$ (c.f. \cite{roberts2016}),
while the disk outflow generates
mostly elements with $A < 140$ (Fig.~\ref{f:histogram_ye}b). The bound tail material
produces more elements with $A > 140$ than the disk material:
although part of it undergoes reprocessing by weak interactions
upon mixing, it is initially more neutron-rich. Given the relative
masses of the ejecta components (Table~\ref{t:models}), yields are dominated by the unbound
tail for $A \gtrsim 90$. The combined abundance distribution 
under-produces elements with $A\lesssim 120$, with minor changes relative
to the pure dynamical ejecta case (c.f.~\cite{roberts2016}).

The total abundance distributions for the set of models that varies the dynamical
ejecta mass are shown in Figure~\ref{f:abundances_combined}b. The change in 
relative contributions from disk, bound tail, and unbound tail lead to
about an order of magnitude variation in the abundances below $A=130$ when
normalizing to the rare-Earth peak. In particular, model Ft0.1 (smallest dynamical ejecta mass) 
is the only case that reaches relative abundance values comparable to Solar at the first 
$r$-process peak ($A=80$), with good overall agreement elsewhere except for some overproduction 
at $A=100$. 
On the other hand, model Ft3.0 (highest dynamical ejecta mass) is hardly different
than a case without disk (Fdyn) for $A \gtrsim 70$.

Regarding abundances for the disk alone, Figure~\ref{f:histogram_ye}b compares the
results for models Fdisk and Fdisk-Y0.1, the latter having $Y_e=0.1$ imposed throughout the
disk at $t=0$. Abundances below $A=130$ are very similar in both cases, with a
ratio of first to second peak abundances that compares well with the Solar System.
This robustness below $A=130$ is consistent with previous nucleosynthesis 
calculations from the disk outflow alone \cite{just2014,wu2016}. The low-$Y_e$
component in model Fdisk-Y0.1 (Fig.~\ref{f:histogram_ye}a) results in a higher
production of elements with $A>130$ than model Fdisk, with an excellent ratio of 
third to second peak abundances and underproduction of rare-Earth elements.
Both models display an abundance excess at $A=100$ relative to the Solar
System distribution. Also, both cases contain an abundance spike at $A=132$
which is related to the convective character of the disk outflow and an
incomplete treatment of nuclear heating in the simulations \cite{wu2016}.

\section{Kilonova Signature}

\begin{figure*}
\centering
\includegraphics*[width=0.565\textwidth]{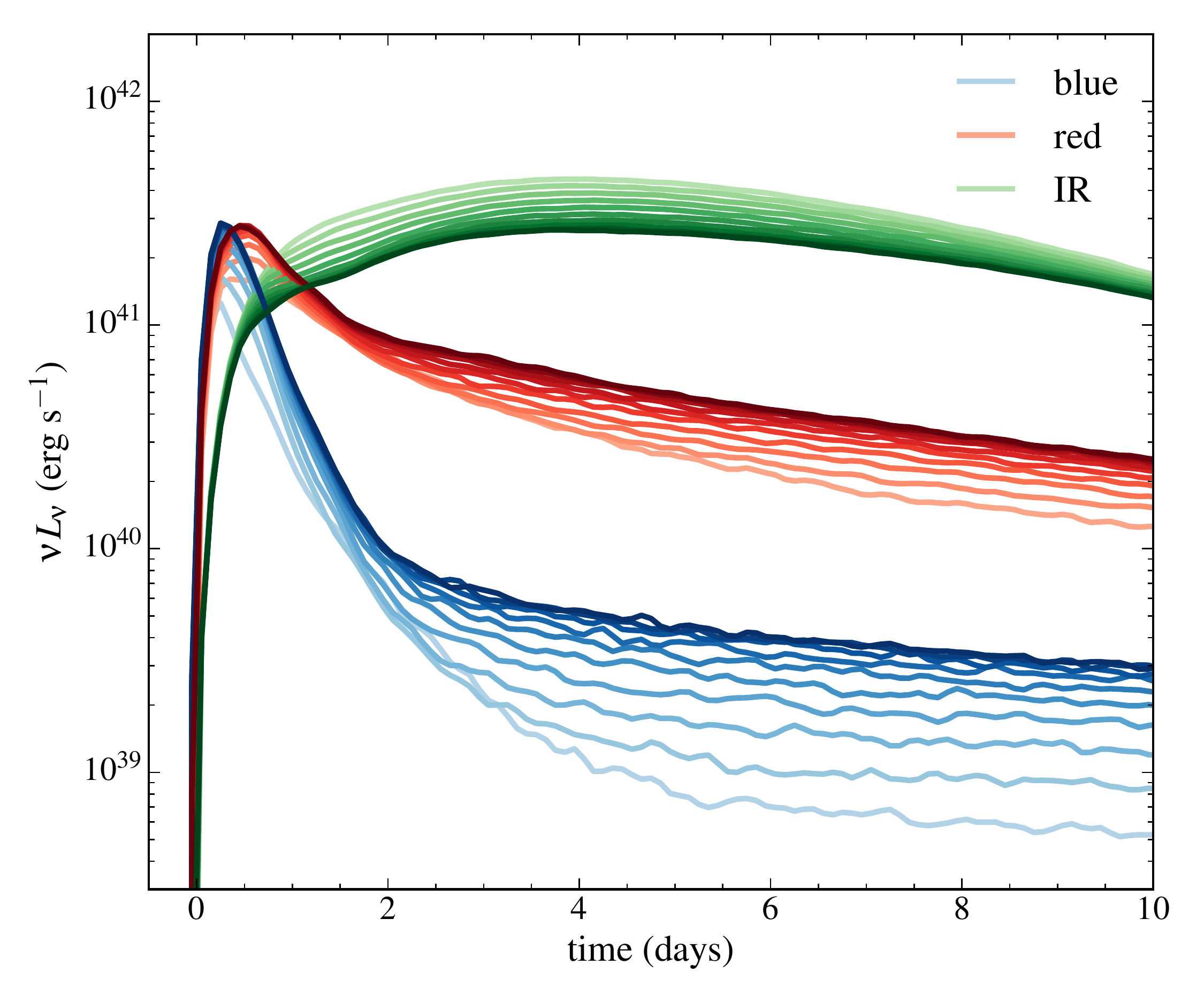}
\includegraphics*[width=0.425\textwidth]{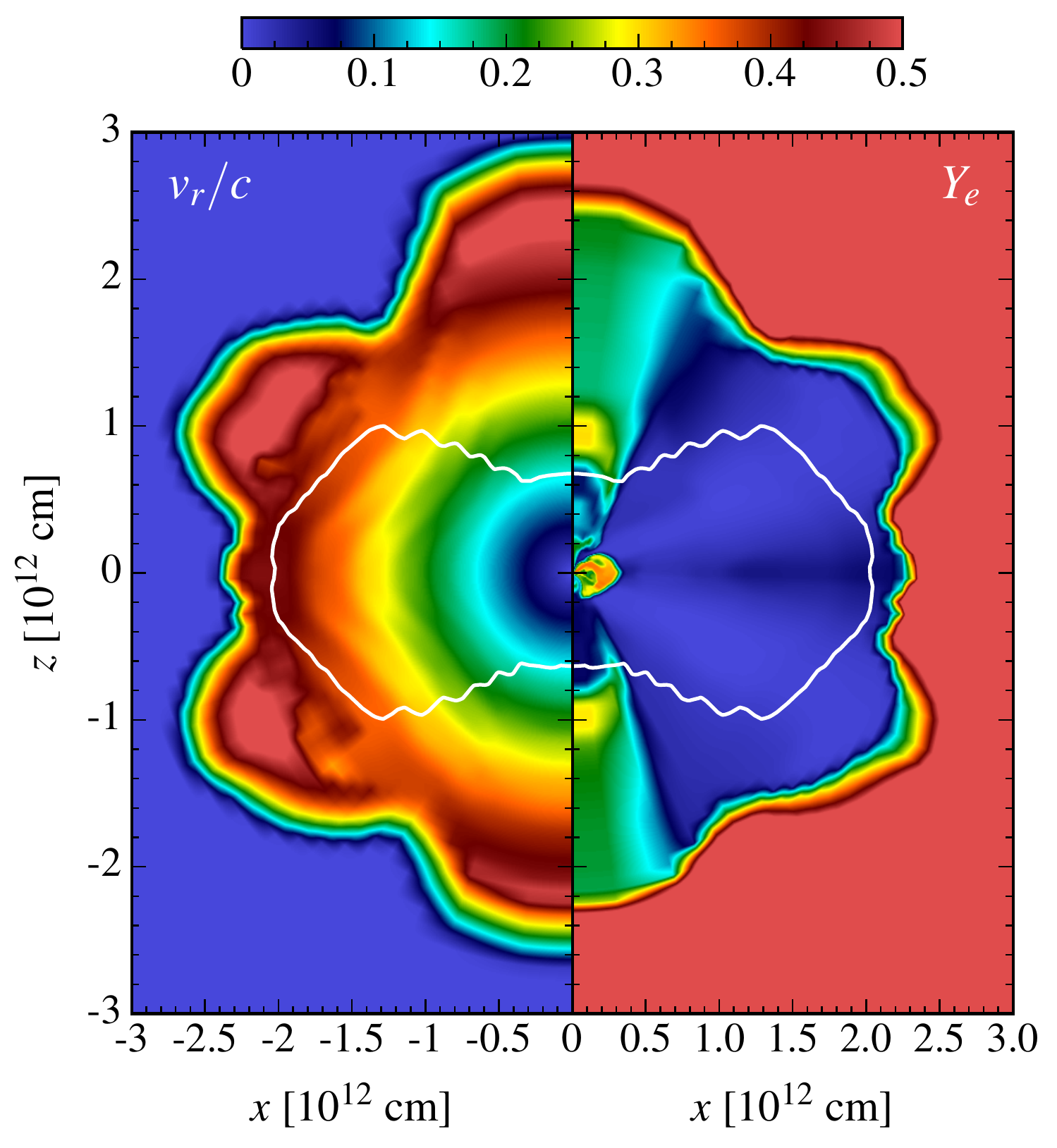}
\centering
\caption{\emph{Left:} Broadband light curves from model F0 in the wave bands $3500-5000\AA$ (blue),
$5000-7000\AA$ (red), and $1-3\mu$m (green). For each color, $10$ viewing angles
equally spaced in their cosine are shown with different shades, spanning the range 
$\theta = 0$ (light, rotation axis) to $\theta=90^\circ$ (dark, equatorial plane).
\emph{Right:} Snapshot of the radial velocity (left panel) and electron fraction (right panel)
in model F0 at time $t=150$~s, when most of the matter distribution has reached homology. This snapshot
is used as input for radiative transfer calculations. The white contour corresponds to a 
density $10^{-6}$~g~cm$^{-3}$.}
\label{f:blue_red_lc}
\end{figure*}

Figure~\ref{f:blue_red_lc}  shows the synthetic kilonova light curves for model
F0. The bolometric luminosity peaks $\approx 5$~days after the merger.  The
emission at the earliest times ($t \lesssim 1$~day) is largely at optical
wavelengths, but the spectral energy distribution (SED) reddens quickly
thereafter. Over the majority of the light curve duration, the radiation
emerges primarily in the near-infrared.  The emission in model F0 is
generated mainly in the dynamical ejecta, which has a mass an order of
magnitude greater than that of the disk wind.  The small amount of $Y_e > 0.25$
(Lanthanide-free) wind ejecta has only a minor impact on the light curves.
Figure~\ref{f:blue_red_lc} shows that this high-$Y_e$ material is surrounded
by lanthanide-rich material in all directions (the high-$Y_e$ `jet' along the
polar axis has very low density).

\begin{figure*}
\centering
\includegraphics*[width=0.9\textwidth]{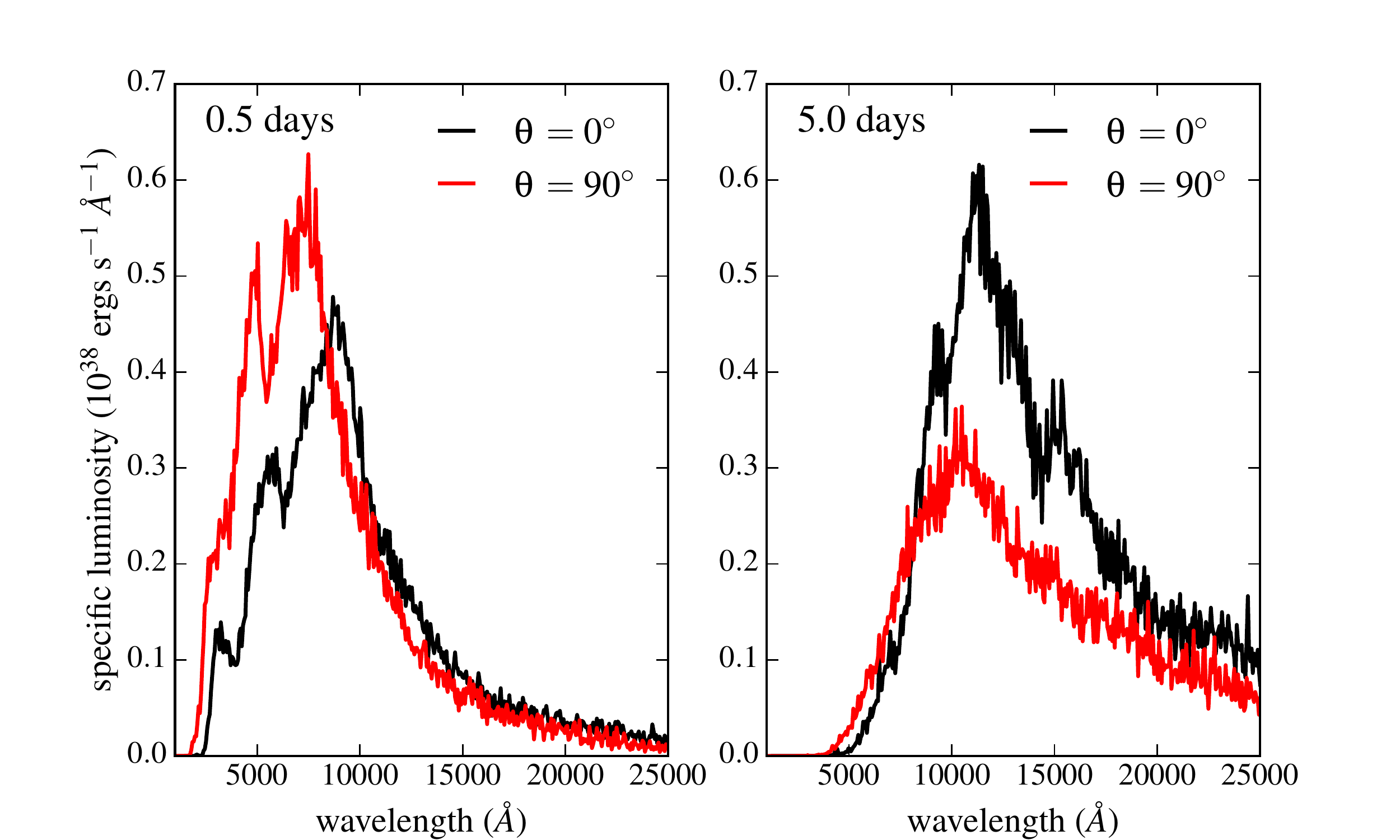}
\centering
\caption{Spectra of model F0 at 0.5 days (left panel) and 5.0 days (right panel) after merger, as observed from both polar ($\theta = 0^\circ$) and equatorial ($\theta = 90^\circ$) viewing angles. The higher equatorial expansion velocity of unbound tail ejecta leads to a greater
blueshift of the spectrum for the $\theta = 90^\circ$ viewing angles.  
The high frequency variations are Monte Carlo noise.}
\label{f:spectra}
\end{figure*}

The spectral evolution of the F0 model (shown in Figure~\ref{f:spectra}) helps
clarify the broadband light curve behavior. At  early times ($t = 0.5$~days)
the entire ejecta on the grid is heated to temperatures  $ \gtrsim 5000$~K and
-- despite the high  opacity from lanthanides  -- the emergent spectra include
an optical component.  By $t \gtrsim 1$~day, the outermost layers of ejecta
have cooled to a photospheric temperature  $\approx 2500$~K, and the SED has
shifted to the near-infrared ($\approx 1~\mu$m), showing little color
evolution thereafter.  This  evolution of the SEDs is similar to that shown for
1D parameterized models by \cite{kasen2013}, who also considered how the early
optical emission is subject to uncertainties in the lanthanide opacities.  In
addition, because the photosphere at early times forms in the outermost layers
of ejecta, the early optical emission may be sensitive to how well the model
resolves the low density surface layers.  

The broadband light curves vary by a factor of $\approx 2$ depending on the
viewing angle.  At most epochs, the kilonova is  bolometrically brightest as viewed from
directions closer the polar axis ($\theta \approx 0^\circ$), due to the fact that the  projected
surface area of the unbound tail is larger by a factor of $\approx 2$ when
viewed face-on (Figure~\ref{f:blue_red_lc}). In the infrared bands, where most of the emission emerges, the
light curves follow this  geometrical orientation dependence. At optical
wavelengths, however, the viewing angle dependence is opposite due to Doppler
shift effects. The observed SED is blueshifted by a characteristic outflow
velocity of $\approx 0.1c$, for polar viewing angles and of $\approx 0.3c$ for
equatorial viewing angles (Figure~\ref{f:blue_red_lc}). Because a greater blueshift  moves 
more of the blue edge of SED  into the optical bands, the optical light curves are always
brightest from equatorial viewing angles (see Figure~\ref{f:spectra}). 

\begin{figure*}
\centering
\includegraphics*[width=\textwidth]{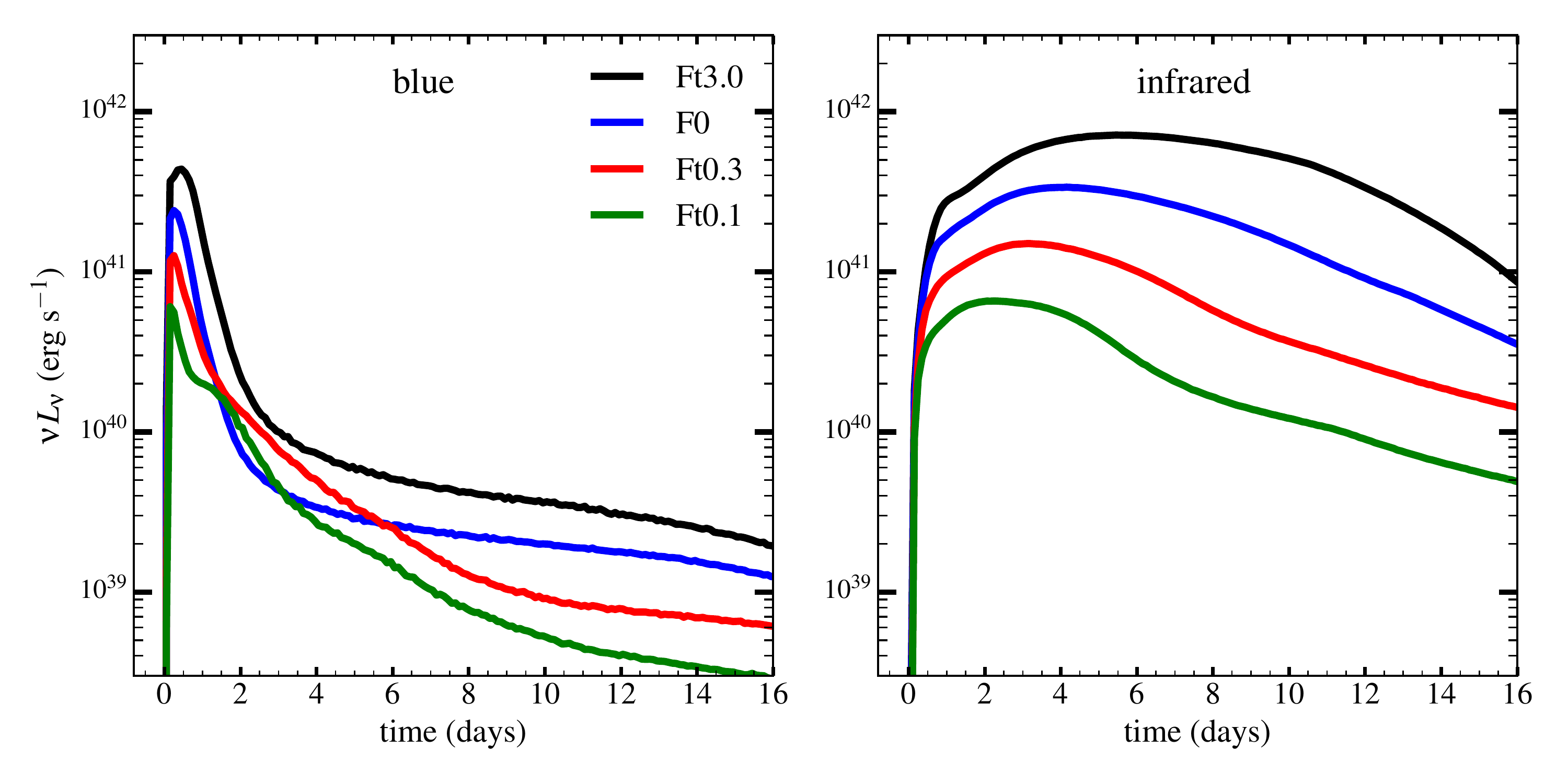}
\centering
\caption{Angle-averaged, broadband light curves from models that scale the dynamical
ejecta mass, as labeled. The left panel shows the blue optical band ($3500-5000\AA$)
and the right panel the infrared band ($1-3\mu$m).}
\label{f:all_lcs}
\end{figure*}

Figure~\ref{f:all_lcs} shows the blue and infrared light curves of the entire set of
models that scale the dynamical ejecta mass, averaged over all viewing angles. 
Peak luminosities vary from
$6\times 10^{40}$~erg~s$^{-1}$ for the dimmest blue optical case to $7\times
10^{41}$~erg~s$^{-1}$ for the brightest infrared luminosity.
The peak time and
luminosity of the infrared emission are monotonic functions of the dynamical
ejecta mass, consistent with the larger amount of radioactive material and
longer diffusion time. The SEDs and orientation effects of these models are
similar to those described for model F0. 

Kasen et al. \cite{Kasen+15} studied the light curves of kilonovae from disk winds, and
showed the degree of blue optical emission depended on the amount of
lanthanide free ($Y_e \gtrsim 0.25$) material ejected. They further showed that  blue
wind emission may be obscured from certain viewing angles if a small mass of
lanthanide-containing dynamical ejecta overlaid the wind. In the models of this
paper, the lanthanide-free disk ejecta is generally obscured from all
viewing angles due to the near complete covering factor of low-$Y_e$ unbound tail
material (Figure~\ref{f:blue_red_lc}). The light curves of these 
models are therefore
mostly insensitive to the amount of lanthanide-free ejecta. As discussed above,
the optical luminosity arises from the  lanthanide-containing dynamical ejecta,
which is hot enough at early times to include some optical component.

\section{Summary}

We have investigated the long-term evolution of the remnant system left behind after
a BH-NS merger, and its electromagnetic and nucleosynthetic signatures. 
Using initial conditions from a GR merger simulation with neutrino transport,
the remnant components are evolved in axisymmetry, accounting for the dominant
nuclear and neutrino source terms. Nucleosynthesis is computed from passive
tracer particles injected in the simulation, and optical/infrared kilonova light curves
are generated using a Monte Carlo radiative transfer code once the ejecta
has reached homologous expansion. We have characterized the dynamics of the
system and its observable signatures when the relative initial masses of disk
and dynamical ejecta are varied. Our main results are the following:
\newline

\noindent 1. -- The presence of the disk prevents fallback material from directly
	        reaching the BH. Instead, fallback can mix efficiently
	        into the disk, increasing the effective disk mass and decreasing the
	        net specific binding energy (Fig.~\ref{f:partial_dens}). The resulting disk 
	        outflow is faster and more neutron-rich as the dynamical ejecta mass increases, at 
	        fixed disk mass (Table~\ref{t:models}).
	        The unbound dynamical ejecta evolves independently of the other components.
                \newline

\noindent 2. -- The total amount of mass ejected in the disk wind outflow increases as the dynamical 
	        ejecta mass increases -- at fixed disk mass -- up to the point where the
	        bound dynamical ejecta mass and disk mass are approximately 
                equal at $t=0$ (Fig.~\ref{f:ejected_fractions}). The mass in Lanthanide- and Actinide-free
		composition also increases up to this point.
                \newline

\noindent 3. -- The composition of the disk outflow is determined primarily by the initial compactness
	        of the disk (mass of the BH divided by the radius of the disk density peak). The
	        compactness sets the strength of the gravitational field, and therefore the
	        disk temperature, the weak interaction timescale, and the beta equilibrium 
		$Y_e$ of the disk. Memory of the initial
	        composition contributes at the $\sim 10\%$ level (Fig.~\ref{f:histogram_ye}).
                \newline
	       
\noindent 4. -- The nucleosynthesis output in the model mapped from the relativistic merger
		simulation (F0) is dominated by the unbound tail 
	        ($A \gtrsim 130$), with
	        a sub-dominant contribution from the disk outflow and turned-around
	        fallback matter at $A\lesssim 130$. Good agreement with a Solar $r$-process distribution
	        is obtained for small dynamical ejecta masses (model Ft0.1), more suitable for
		small BH masses  (Fig.~\ref{f:abundances_combined}). Excess production of elements around $A=100$ and
	        at $A=132$ is obtained.
                \newline

\noindent 5. -- The kilonova signature produces optical emission for the first day after merger, 
                then evolves to the infrared. The peak luminosities are of order
	        $10^{41}$~erg~s$^{-1}$, with a variation of a factor
	        $\sim 2$ obtained for different viewing angles (Fig.~\ref{f:blue_red_lc}).
		The optical lightcurves are brightest when the system is viewed 
		edge-on, due to Doppler shift effects.
	 	The peak optical and infrared luminosities, as well as the transient
	   	duration, are monotonically increasing functions of the ratio of
	        initial dynamical ejecta to disk masses (Fig.~\ref{f:all_lcs}),
	        or equivalently of the total ejected mass. The \emph{detailed properties} of the
                early optical emission are sensitive to the outer edges of the disk and
	        dynamical ejecta, and thus will depend on how well resolved are
	        low-density regions in simulations. Nevertheless we consider the general properties
	        of our results are robust.
                \newline

Our results on the long-term ejecta dynamics of the model with physical parameter (F0) 
differ from our previous study, which employed
initial conditions from a Newtonian merger simulation (Fern\'andez et al. 2015; Ref. \cite{FQSKR2015}). 
Results from that work showed only minor mixing between \emph{fallback} and disk material in the outer
parts of the disk outflow (see e.g. Figure~3 in that paper and the insensitivity of the electron 
fraction of the disk outflow to the presence or absence of dynamical ejecta in their
Table 2). This difference can be attributed to the different relative
masses of disk and bound dynamical ejecta, and on the way the initial conditions are obtained
here. In our model F0,
the ratio of fallback-to-disk masses is 0.038/0.060, whereas in the baseline
model C2d of Ref. \cite{FQSKR2015} this ratio is 0.02/0.20. The bound dynamical ejecta reaches a peak 
density a factor of $\sim 16$ lower in C2d than in our model F0, hence it is not 
able to penetrate the disk and mix in efficiently (in contrast to our Figure~\ref{f:mixing}). 
A similar behavior is observed in model Ft0.1, for which the bulk of the disk remains unaffected 
by fallback material, which only mixes in on its periphery (Figure~\ref{f:mixing_x01}). Still, model Ft0.1 has a
mean $Y_e$ of the disk outflow (0.30) that is lower by $14\%$ relative to the model without
dynamical ejecta (Fdisk, 0.35), whereas the corresponding fractional decrease in the models of Ref. \cite{FQSKR2015} 
is about half (0.29 to 0.27, respectively). This lower mixing efficiency could result from
 a number of factors, including a higher concentration of dynamical ejecta in the equatorial
plane in the general relativistic simulation, and on the fact that the disk and dynamical 
ejecta used by Ref. \cite{FQSKR2015} are not computed separately, hence some mixing is 
already built into it (the mean $Y_e$ of the resulting outflow compares favorably to 
the average between our models Ft0.1 and Ft0.3, which bracket the initial 
fallback-to-disk mass ratio in model C2d of \cite{FQSKR2015}).

An important observational implication of our results is that a substantial
amount of blue optical emission can still be generated by the Lanthanide-rich
ejecta at early times when the temperatures are high. 
The duration of this signal is $\lesssim 1$~day,
requiring a shorter cadence for its detection. 
A similar early optical component was seen in the Lanthanide-rich wind models
of \cite{Kasen+15}. The robustness of the predicted optical emission is
subject to uncertainties in the incomplete Lanthanide atomic data
\cite{kasen2013}; further radiative transfer calculations and atomic structure
models are needed to fully assess the early time kilonova colors.
A peak optical luminosity of $10^{41}$~erg~s corresponds to an absolute magnitude
$M_B \simeq -14$ (apparent magnitude $m_B\simeq 24.2$ at $400$~Mpc). 

The general properties of our light curves compare favorably to previous work
on kilonova emission from BH-NS mergers. Roberts~et~al. \cite{Roberts+11}
showed that the asymmetrical distribution of dynamical ejecta from NS-NS
mergers leads to viewing angle dependences with the brightness varying by a
factor of $\approx 2$. A similar magnitude of viewing angle effects was found
by \cite{grossman2014}.
Kyutoku et al. \cite{kyutoku2013} emphasized the
asymmetries intrinsic in the BH-NS merger dynamical ejecta due to the dominance
of tidal forces on the ejection. They also estimated a factor of a few variation
in the kilonova light curve properties at peak as a function of viewing angle. Tanaka et al. \cite{tanaka2014}
carried out multi-dimensional Monte Carlo radiative transfer, and showed that in addition to
variations in the line of sight, the asymmetries in the dynamical ejecta cause
the emission to be systematically bluer than in the case of NS-NS mergers.
In contrast to their results, however, we find that asymmetries due to the
line of sight persist beyond 10 days, particularly in the optical band.
Finally, Kawaguchi et al. \cite{kawaguchi2016} conducted a parameter space study of kilonova
properties from BH-NS mergers by developing a fitting formula for the dynamical ejecta mass
and for the kilonova light curves as a function of binary mass ratio, BH spin,
and NS size (EOS). Our scaling of dynamical ejecta mass at fixed disk
mass is a reasonable approximation to the variation with BH spin, for fixed EOS.

Our results are consistent with our previous work \cite{FQSKR2015} when it comes to
the source of late-time accretion onto the compact object. Figure~\ref{f:mass_fluxes_tidal}
shows that accretion of `fallback' material follows the same time
dependence as accretion of disk material, with the slope set by disk physics
rather than the usual $\sim t^{-5/3}$ obtained when assuming a flat energy distribution
of material in Keplerian orbits \cite{rees1988,rosswog07,metzger2009_fallback}. 
A reliable prediction of this time-dependence
requires carrying out disk simulations in magnetohydrodynamics (MHD).
For instance, the normalization and time-dependence of the disk accretion that we
obtain in our models (e.g., Figure~11 of \cite{FQSKR2015}) would imply insufficient accretion energy
as required to power the X-ray emission from GRB 130603B at a time of $\sim 1$~day \cite{fong2014}.
However, relatively small changes in the time exponent could bring the energy production into agreement.
Whether the X-rays themselves contribute to the powering of the kilonova, as proposed by 
\cite{kisaka2016}, is a separate question that depends on the degree of beaming of the highly 
super-Eddington emission obtained \cite{metzger2016}. Pursuit of these questions is beyond
the scope of this study.

Improving the reliability of these calculations would require the following: (1) general-relativistic
merger simulations that employ a larger computational domain, so that initial conditions for both disk and 
dynamical ejecta are obtained self-consistently, and which use higher resolution so that
low-density regions are more reliably evolved (2) long-term general-relativistic, MHD
simulations in 3D to evolve both the disk and dynamical ejecta on a fixed metric and with a more
realistic geometry, including the 
dominant neutrino and nuclear source terms, and (3) more reliable opacities for $r$-process elements, 
which when combined with coupling to the actual composition of the outflow (instead of just $Y_e$), 
would yield spectral predictions alongside broadband light curves. The ejecta and disk properties
can be sensitive to parameter dependencies other than just the mass ratio, such as the
rotation of the neutron star (e.g., \cite{dietrich_2017}) or effects due to non-circularity
of the binary (such as e.g., formation in dynamical captures, \cite{paschalidis_2015}).
The development of open-source radiative transfer codes would also allow end-to-end
studies to be carried out by a wider community; to our knowledge, the only open-source
tool available to date is SNEC \cite{morozova_2015}.
Further improvements in the calculations will reveal whether the trends found
here are robust.

\ack
We thank Kenta Hotokezaka, Masaomi Tanaka, Kunihito Ioka, and Shinya Wanajo for useful discussions.
We also thank the anonymous referee for helpful comments that improved the presentation of this paper.
RF acknowledges support from the University of California Office of the President,
from NSF grant AST-1206097, and from the Faculty of Science at the University
of Alberta.
Support for this work was provided by NASA through Einstein Postdoctoral Fellowship 
grant numbered PF4-150122 (FF) awarded by the Chandra X-ray Center, 
which is operated by the Smithsonian Astrophysical Observatory for NASA under contract
NAS8-03060.
DK is supported in part by a Department of Energy Office of Nuclear
Physics Early Career Award, and by the Director, Office of Energy
Research, Office of High Energy and Nuclear Physics, Divisions of
Nuclear Physics, of the U.S. Department of Energy under Contract No.
DE-AC02-05CH11231.
The software used in this work was in part developed by the DOE NNSA-ASC OASCR Flash Center at the
University of Chicago.
This research used resources of the National Energy Research Scientific Computing
Center (NERSC), which is supported by the Office of Science of the U.S. Department of Energy
under Contract No. DE-AC02-05CH11231. Computations were performed at
\emph{Edison} (repositories m1186 and m2058).
This research also used the \emph{Savio} computational cluster resource provided by the
Berkeley Research Computing program at the University of California, Berkeley
(supported by the UC Berkeley Chancellor, Vice Chancellor of Research, and
Office of the CIO). We also thank the hospitality of the Yukawa Institute for
Theoretical Physics through the workshop Nuclear Physics, Compact Stars, and Compact
Star Mergers 2016, during which part of this work was carried out.

\section*{References}

\bibliographystyle{iopart-num}
\bibliography{rodrigo,skynet}

\end{document}